\makeatletter
\declare@file@substitution{revtex4-1.cls}{revtex4-2.cls}
\makeatother

\documentclass[twocolumn]{aastex631}

\usepackage[utf8]{inputenc}
\usepackage[T1]{fontenc}
\usepackage{amsmath,amsfonts,amssymb,bm,graphicx,booktabs,xcolor}
\usepackage[all]{hypcap}
\hypersetup{
    colorlinks=true,
    linkcolor=magenta,
    citecolor=blue,
    filecolor=magenta,
    urlcolor=blue,
}

\newcommand{\ba}{\begin{eqnarray}}
\newcommand{\ea}{\end{eqnarray}}

\newcommand{\tc}{T_c}
\newcommand{\af}{a_{\rm f}}
\newcommand{\nf}{n_{\rm f}}
\newcommand{\omeff}{\omega_{\rm eff}}
\newcommand{\taum}{\tau_M}
\newcommand{\qeff}{Q_{\rm eff}}

\shorttitle{Thermally Regulated Tidal Migration}
\shortauthors{Petrovich \& Bhaskar}

\begin{document}

\title{Thermally Regulated Viscoelastic Tidal Migration of Eccentric Planets}

\author[0000-0003-0412-9314]{Cristobal Petrovich}
\affiliation{Department of Astronomy, Indiana University, Bloomington, IN, USA}

\author[0000-0002-5181-0463]{Hareesh Gautham Bhaskar}
\affiliation{Department of Astronomy, Indiana University, Bloomington, IN, USA}

\begin{abstract}
A growing population of short-period Neptune-like planets has nonzero eccentricities and, in some cases, large stellar obliquities, suggestive of high-eccentricity migration. Modeling this evolution requires a prescription for tidal dissipation, which in Neptune-like planets may be dominated by rocky or icy cores rather than by extended gaseous envelopes as often assumed for gas giants. We study the coupled orbital and thermal evolution of eccentric Neptune-like planets whose tidal dissipation is controlled by a viscoelastic Maxwell core. We compute the full harmonic tidal response, follow the equilibrium spin state, and evolve the core temperature; the changing thermal state feeds back on the orbit by modifying the viscosity and hence the frequency-dependent tidal response. We find that the coupled evolution is self-regulated: tidal heating initially drives cold, weakly dissipative cores toward efficient dissipation, but the resulting temperature increase lowers the viscosity, shifts the tidal response toward the fluid-like side of the Maxwell peak, and slows the subsequent migration. This feedback drives the system toward a quasi-steady heating-cooling balance, produces long-lived eccentric phases, and weakens the otherwise steep dependence of circularization time on final orbital distance. Thermally regulated tides therefore offer a natural explanation for how observed hot Neptunes can remain eccentric across a broad range of orbital distances, where a single fixed tidal efficiency would either erase eccentricity too efficiently close in or fail to damp it farther out.
\end{abstract}


\section{Introduction}\label{sec:intro}

The close-in Neptune population contains several clues to its dynamical history. Neptune-like planets are depleted at the shortest orbital periods, forming the Neptunian desert, while an overdensity of intermediate-size planets appears near its edge at periods of a few days \citep[e.g.,][]{Mazeh2016,Bourrier2023,CastroGonzalez2024}. This ``Neptunian ridge'' lies near the hot-Jupiter pileup and has been interpreted as evidence that some Neptunes reach short-period orbits through high-eccentricity migration (HEM), possibly late enough to retain their envelopes against irradiation-driven atmospheric escape \citep{Bourrier2018,Bourrier2023,Correia2020,CastroGonzalez2024,CastroGonzalez2026,Lu2025,YuDai2025,LoRusso2026}. Significant spin--orbit misalignments among several close-in Neptunes and sub-Saturns provide additional evidence for dynamically excited migration \citep[e.g.,][]{Albrecht2022,Attia2023,Rubenzahl2024,Espinoza2024,Espinoza2026}.

A particularly intriguing feature is the persistence of eccentricity among close-in Neptune-mass planets. \citet{Correia2020} highlighted that many warm Neptunes retain eccentricities of order $e\sim0.1$--$0.2$ despite experiencing strong tidal forcing. More recently, \citet{Wang2026} identified a population of eccentric and spin--orbit misaligned sub-Saturns around single stars, a combination that is comparatively rare among hot Jupiters. In Appendix~\ref{sec:appendix_observations}, we further show tentative evidence that the eccentric fraction of isolated hot Neptunes depends only weakly on the inferred circularization radius $a_f=a(1-e^2)$, in contrast to the much steeper decline observed among hot Jupiters.

These observations are difficult to reconcile with standard equilibrium-tide prescriptions having fixed tidal efficiency. Along a constant-angular-momentum migration track, constant time-lag (CTL) and constant-$Q$ models predict eccentricity damping times that depend steeply on orbital separation, approximately as $t_e\propto a_f^8$ for CTL models \cite{Hut81}. While such behavior is broadly consistent with close-in Jupiters \citep{jackson2008,Hansen2010}, Neptune-mass planets differ in an important respect: a substantial fraction of their mass resides in dense rocky or icy interiors, where tidal dissipation may be dominated by a viscoelastic core. Viscoelastic core dissipation has previously been explored in the context of HEM of giant planets \citep{StorchLai2014,correia2014}, but without coupling the tidal response to the planet's thermal evolution.

Because the viscosity of rocky and icy materials depends strongly on temperature, tidal heating can modify the dissipation rate itself, producing thermal--viscous feedback. Similar feedback has been studied for rocky planets, satellites, exomoons, and lava worlds, typically in low- to moderate-eccentricity settings where the tide is described by a small number of forcing frequencies \citep[e.g.,][]{Ojakangas1986,Segatz1988,Henning2009,Meyer2010,Zahnle2015,DobosTurner2015_exomoons,Driscoll2015,Tian2017,Herath2024,Herath2026}. In HEM, by contrast, the planet passes through very high eccentricities, where the forcing is spread over many Hansen/Kaula harmonics whose frequencies evolve with both eccentricity and spin.

Here we study this feedback for eccentric Neptune-like planets. We compute the full eccentric tidal spectrum, couple the resulting dissipation to the thermal evolution of a Maxwell core, and show that the tidal efficiency self-regulates as the planet migrates.

\begin{figure*}[t!]
\centering
\includegraphics[scale=0.1]{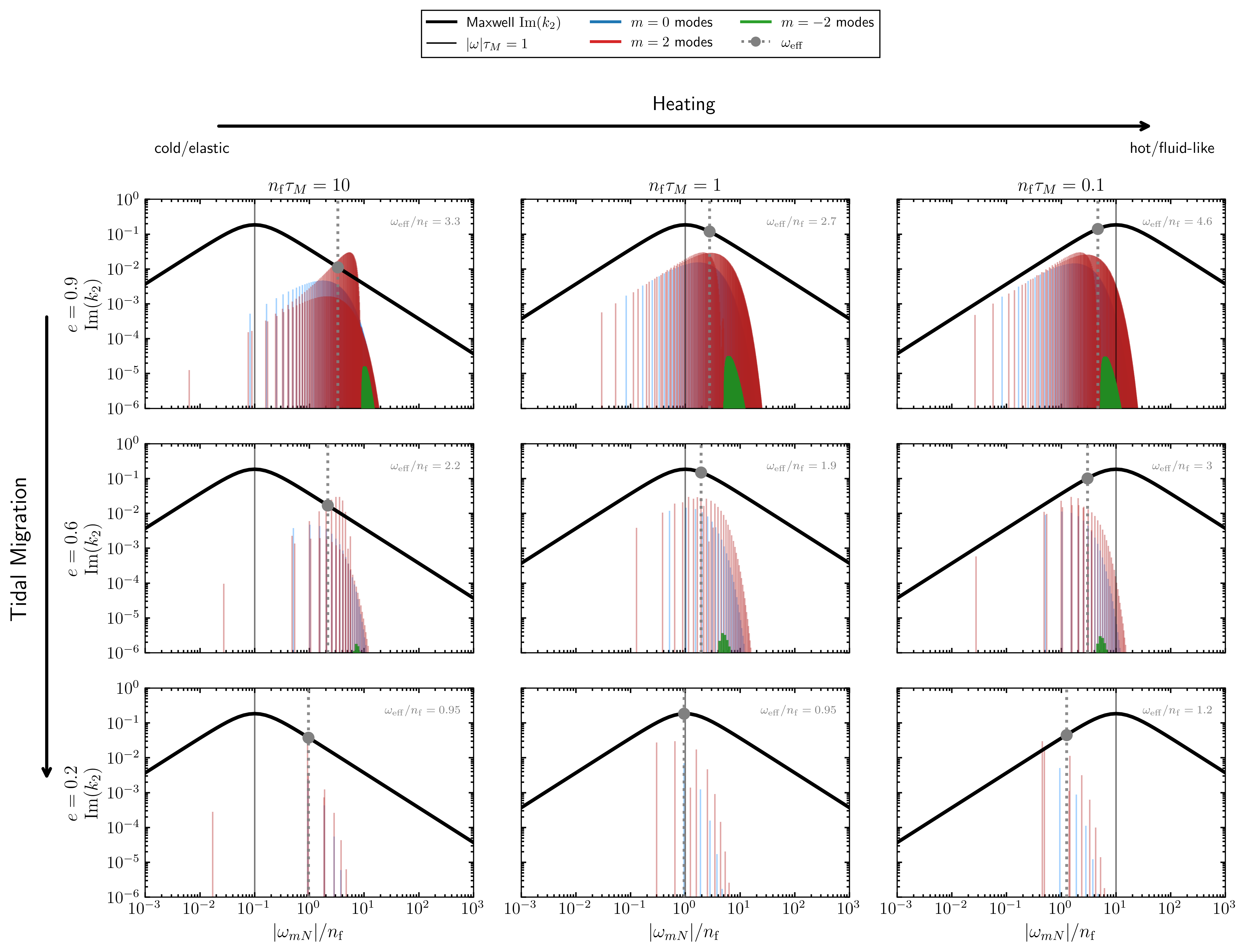}
\caption{
Maxwell response and eccentric tidal forcing spectrum. Rows show decreasing eccentricity, corresponding to tidal migration along a constant-angular-momentum track; columns show decreasing $\nf\taum$, corresponding to heating and a shorter tidal relaxation time. Black curves show the Maxwell response $\mathrm{Im}(k_2)$, and solid vertical black lines mark the Maxwell peak, $|\omega|\taum=1$. Colored vertical lines show the relative modal heating weights $(|\omega_{mN}|/\nf)\mathcal{F}_{mN}$ for $m=0,\pm2$, where $\mathcal{F}_{mN}=|W_{2m}F_{mN}|^2$. Frequencies are measured in the planetary rotating frame and normalized by the final circular mean motion $\nf$. Gray dotted lines and circles mark the dissipation-weighted effective forcing frequency $\omeff$. At high eccentricity, the heating is produced by the overlap of many high-order harmonics with the viscoelastic response, rather than by a single forcing frequency.
}
\label{fig:modal_spectrum}
\end{figure*}

\section{Eccentric tidal forcing and viscoelastic response}\label{sec:formalism}

We consider tides raised by a star of mass \(M_\star\) on a Neptune-like planet whose dissipation is dominated by a rocky or icy core of radius \(R_c\) and mass $M_c$. We restrict attention to the planar problem and neglect planetary obliquity, so that the planet's spin is aligned with the orbital angular momentum. For the quadrupolar tide considered here, \(m=0,\pm2\), and throughout this paper \(\sum_{m,N}\) denotes a sum over these three values of \(m\) and over all integer \(N\). The quadrupolar tidal heating rate \citep[e.g.,][]{correia2014,StorchLai2014} is written as
\begin{equation}
\dot E_{\rm tide}
=
\frac{5}{4\pi}
\frac{G M_\star^2 R_c^5}{a^6}
\sum_{m,N}
\mathcal{F}_{mN}(e)\,
\omega_{mN}\,
\mathrm{Im}\!\left[k_2(\omega_{mN})\right],
\label{eq:edot}
\end{equation}
where
\begin{equation}
\mathcal{F}_{mN}(e)=\left|W_{2m}F_{mN}(e)\right|^2,
\label{eq:fmn}
\end{equation}
and
\begin{equation}
\omega_{mN}=Nn-m\Omega_s
\label{eq:omega_mn}
\end{equation}
is the forcing frequency in the rotating frame of the planet. The Hansen coefficients are
\begin{equation}
F_{mN}(e)=
\left<
\left(\frac{a}{r}\right)^3 e^{imf} e^{-iNM}
\right>,
\label{eq:hansen}
\end{equation}
where $f$ and $M$ are the true and mean anomalies, respectively, and the brackets denote an orbital average. The normalization constants satisfy
$W_{20}^2=\pi/5$
and
$W_{2,\pm2}^2=3\pi/10$.
In practice, we truncate the harmonic expansion at sufficiently large $|N|$ (typically $|N|\sim200$--$500$) to ensure convergence for the eccentricities considered.

The core is modeled as a self-gravitating Maxwell body. For a forcing frequency $\omega$, we write the dissipative part of the quadrupolar Love number as \citep{correia_valente2022}:
\begin{equation}
\mathrm{Im}\!\left[k_2(\omega)\right]
=
(k_f-k_e)
\frac{\omega\taum}{1+(\omega\taum)^2},
\label{eq:maxwell}
\end{equation}
where \(k_f\) and \(k_e\) are the fluid and elastic Love numbers, respectively. The fluid value \(k_f\) is the zero-frequency limit of the real Love-number response, while \(k_e\) is the high-frequency elastic limit; their difference \(k_f-k_e\) sets the amplitude of the dissipative response. For a homogeneous self-gravitating Maxwell body, the effective Maxwell relaxation time of the tidal response is 
\begin{equation}
\taum=\tau_e+\tau_v,
\qquad
\tau_e=\frac{\eta}{\mu},
\qquad
\tau_v=\frac{38\pi\eta R_c^4}{3GM_c^2},
\label{eq:maxwell_times}
\end{equation}
where $\eta$ and $\mu$ are the viscosity and shear modulus of the dissipating core. Here $\tau_e$ is the material Maxwell time, while $\tau_v$ is the gravity-controlled viscous relaxation time. For this Maxwell model, the Love-number contrast is related to the two relaxation times by
\begin{equation}
k_f-k_e
=
k_f
\frac{\tau_v}{\taum}
=
k_f
\frac{\mu}{\mu+\mu_{\rm grav}},
\label{eq:kcontrast}
\end{equation}
where
\begin{equation}
\mu_{\rm grav}\equiv
\frac{3GM_c^2}{38\pi R_c^4}
\label{eq:mugrav}
\end{equation}
is the self-gravity scale for tidal deformation, sometimes called the gravitational rigidity. Dissipation peaks near $|\omega|\taum\simeq1$, separating the elastic regime, $|\omega|\taum\gg1$, from the fluid-like regime, $|\omega|\taum\ll1$.

We assume that the spin angular momentum of the planet is small compared to the orbital angular momentum and neglect tides raised on the star. In this limit, tides raised on the planet primarily remove orbital energy, while the orbital angular momentum is approximately conserved. The migration therefore follows
\begin{equation}
a(1-e^2)=\af,
\label{eq:constant_J}
\end{equation}
where $\af$ is the final circularized semimajor axis. Equivalently,
\begin{equation}
n(e)=\nf(1-e^2)^{3/2},
\qquad
\nf=\left(\frac{GM_\star}{\af^3}\right)^{1/2}.
\end{equation}
For fixed $\af$, the dimensionless parameter $\nf\taum$ determines which part of the eccentric tidal spectrum overlaps the Maxwell peak, while $k_f-k_e$ sets the overall strength of the dissipative response.

We further assume that the planetary spin rapidly evolves toward a stable zero-torque tidal equilibrium,
\begin{equation}
\Omega_s=f_{\rm eq}(e,\nf\taum)\,n .
\end{equation}
The function $f_{\rm eq}$ is obtained by requiring the net tidal torque to vanish, following the procedure described in Appendix~\ref{sec:app_spin}. When multiple zero-torque solutions exist, we select the highest-frequency stable branch. In the weak-friction limit, $\nf\taum\ll1$, this prescription recovers the classical pseudo-synchronous result of \citet{Hut81},
\begin{equation}
f_{\rm eq}(e,\nf\taum\ll1)
\rightarrow
f_{\rm Hut}(e)
\equiv
\frac{f_2(e)}
{f_5(e)(1-e^2)^{3/2}},
\label{eq:fps}
\end{equation}
where the eccentricity polynomials $f_2$ and $f_5$ are defined in Appendix~\ref{secc:app_quasi_steady}.

To characterize the eccentric tidal spectrum, we define the dissipation-weighted effective forcing frequency
\begin{equation}
\omeff=
\frac{
\sum_{m,N}
\mathcal{F}_{mN}|\omega_{mN}|\,\omega_{mN}
\mathrm{Im}[k_2(\omega_{mN})]
}
{
\sum_{m,N}
\mathcal{F}_{mN}\omega_{mN}
\mathrm{Im}[k_2(\omega_{mN})]
}.
\label{eq:omega_eff}
\end{equation}
Since $\omega\,\mathrm{Im}[k_2(\omega)]>0$, this quantity is positive definite and measures the characteristic absolute forcing frequency of the modes that contribute to the heat.

We use the effective forcing frequency to define a representative tidal quality factor normalized by the Love-number contrast,
\begin{equation}
Q_{\rm eff}
\equiv
\frac{k_f-k_e}
{\left|\mathrm{Im}\left[k_2(\omega_{\rm eff})\right]\right|}.
\label{eq:qeff}
\end{equation}
For the Maxwell response in Equation~\eqref{eq:maxwell},
\begin{equation}
Q_{\rm eff}
=
\frac{1+(\omega_{\rm eff}\taum)^2}
{\omega_{\rm eff}\taum}.
\end{equation}
This definition assigns the broadband eccentric tidal spectrum a single
representative quality factor evaluated at the dissipation-weighted forcing
frequency. It is intended to facilitate comparison with constant-\(Q\) tidal
models; the tidal heating and torque are still computed from the full harmonic
sum rather than from a prescribed constant \(Q\).

\subsection{The eccentric tidal spectrum}

The definitions above reduce the problem to the overlap between the eccentric tidal forcing spectrum and the Maxwell response of the core. Figure~\ref{fig:modal_spectrum} illustrates this overlap for three eccentricities and three values of $\nf\taum$. Each vertical line represents a harmonic contribution to the heating, grouped by azimuthal order $m$, while the black curve shows the Maxwell response. The gray vertical line marks $|\omega|\taum=1$, where dissipation peaks, and the gray point marks the dissipation-weighted effective forcing frequency $\omeff$.

At high eccentricity the tidal potential is spread over many Hansen/Kaula harmonics. The non-axisymmetric components are shifted by the spin through
\begin{equation}
\omega_{mN}=Nn-m\Omega_s .
\end{equation}
Because the zero-torque spin is rapid at high eccentricity, the corotation condition for the $m=2$ terms occurs at
\begin{equation}
N\simeq 2\frac{\Omega_s}{n}=2f_{\rm eq}.
\end{equation}
For example, at $e=0.9$ the pseudo-synchronous spin is large, $f_{\rm eq}\sim f_{\rm Hut}\simeq 36$, so corotation lies near $N\sim70$. Thus the dominant non-axisymmetric harmonics are high-order modes, and the heating cannot be represented by a single frequency comparable to the mean motion.

The figure also shows how migration moves the planet across the spectrum. Along a constant-angular-momentum track, decreasing eccentricity narrows the Hansen spectrum and shifts the dominant harmonics to lower $N$. This is visible from the upper to lower panels: as $e$ decreases from 0.9 to 0.2, the mode distribution contracts and $\omeff$ decreases from several times $\nf$ to order $\nf$. This occurs even though the instantaneous mean motion increases during circularization, since $n=\nf(1-e^2)^{3/2}$ grows as $e$ declines. The relevant forcing frequency for dissipation is therefore set by the evolving harmonic spectrum and spin state, not simply by $n$.

The horizontal direction in Figure~\ref{fig:modal_spectrum} corresponds to thermal evolution at fixed orbit. As the core heats, $\taum$ decreases, moving the Maxwell peak to larger $|\omega|/\nf$ in the plotted units. A planet can therefore pass from the elastic regime, through the Maxwell peak, and into the fluid-like regime as its temperature rises. The tidal evolution is controlled by this two-dimensional motion: orbital circularization changes the spectrum itself, while thermal evolution changes where the Maxwell response overlaps that spectrum.

\begin{figure*}
\centering
\includegraphics[scale=0.085]{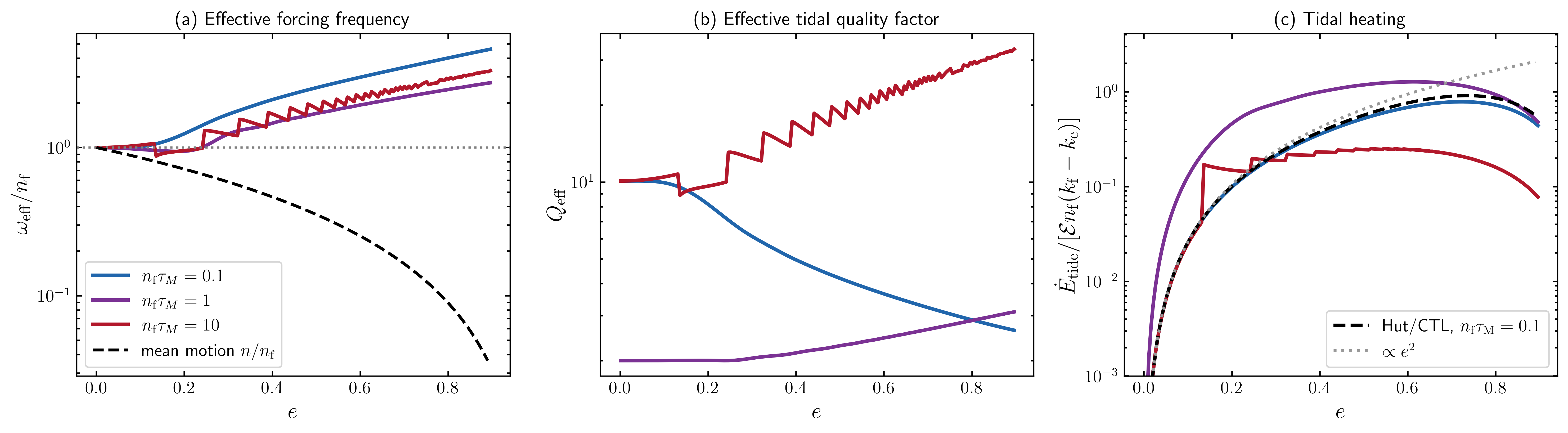}
\caption{
Effective forcing frequency, tidal quality factor, and heating along constant-angular-momentum migration tracks. The calculation uses the full eccentric forcing spectrum for fixed values of $\nf\taum$. Panel (a) shows that the dissipation-weighted forcing frequency $\omeff$ can differ substantially from both $n$ and $\nf$. Panel (b) shows the corresponding representative quality factor $Q_{\rm eff}$, which varies as the eccentric spectrum moves relative to the Maxwell peak. Panel (c) shows the dimensionless tidal heating rate. Heating peaks at intermediate eccentricity because the eccentric forcing weakens at low $e$, while the large semimajor axis along a constant-angular-momentum track suppresses the tidal amplitude as $e\rightarrow1$. The dashed curve shows the weak-friction Hut limit for $\nf\taum=0.1$, and the dotted curve shows the leading $e^2$ scaling.
}
\label{fig:q_heat}
\end{figure*}

\subsection{Heating along a constant-angular-momentum track}

Before coupling the tides to thermal evolution, we first examine the heating produced by a fixed Maxwell rheology. We hold $\taum$ constant, label each model by $\nf\taum$, and follow a planet along a constant-angular-momentum track, $a(1-e^2)=\af$. This isolates how the changing eccentricity modifies the harmonic spectrum and its overlap with the Maxwell response.

We normalize the heating by
\begin{equation}
\mathcal{E}
\equiv
\frac{5}{4\pi}
\frac{GM_\star^2R_c^5}{\af^6},
\label{eq:Escale}
\end{equation}
so that $\dot E_{\rm tide}$ is measured in units of $\mathcal{E}\nf$.

Figure~\ref{fig:q_heat} summarizes the result. Panel~(a) shows the dissipation-weighted forcing frequency $\omeff$. Because the heating receives contributions from many harmonics, $\omeff$ is not generally comparable to the instantaneous mean motion. At high eccentricity it can exceed $\nf$ by several factors, while the instantaneous mean motion $n=\nf(1-e^2)^{3/2}$ becomes small along the constant-angular-momentum track. Thus the characteristic frequency relevant for dissipation is set by the eccentric spectrum and spin state, not by $n$ alone. The choppy structure for $\nf\taum=10$ reflects jumps between neighboring stable spin-orbit resonances in the solid-response regime, which change the rotating-frame frequencies of the dominant harmonics.

Panel~(b) shows the corresponding $Q_{\rm eff}$. The quality factor is not fixed: it decreases when the dominant harmonics approach the Maxwell peak and increases when they lie farther into either the elastic or fluid-like regime. The same eccentricity can therefore correspond to different dissipation efficiencies depending on $\nf\taum$.

Panel~(c) shows the resulting heating rate. In all cases, the maximum occurs at intermediate eccentricity. At low $e$, the eccentric tidal forcing is weak. At very high $e$, the constant-angular-momentum relation implies a large semimajor axis, reducing the overall amplitude through the $a^{-6}$ factor in Equation~\eqref{eq:edot}. The peak shifts with rheology because changing $\nf\taum$ changes where the Maxwell response overlaps the eccentric spectrum.

For comparison, panel~(c) also shows the weak-friction limit of \citet{Hut81} along the same constant-angular-momentum track,
\ba
\frac{\dot E_{\rm tide}}{\mathcal{E}n_f(k_f-k_e)}
&=&
\frac{12\pi}{5}
\left(\taum n_f\right)
\left(1-e^2\right)^{3/2}
F_{\rm Hut}(e)
\label{eq:hut_dimensionless}\\
&=&
\frac{42\pi}{5}
\left(\taum n_f\right)e^2
+\mathcal{O}(e^4),
\label{eq:hut_e2}
\ea
where the plotted curve uses $\nf\taum=0.1$. This expression tracks the blue Maxwell curve because that model is close to the weak-friction regime. The leading $e^2$ approximation captures only the low-eccentricity behavior; it misses the turnover of the full Hut expression and cannot describe the additional shifts produced by frequency-dependent viscoelastic dissipation.

Figure~\ref{fig:q_heat} is therefore an instantaneous map for fixed rheology. The next step is to allow $\taum$ to evolve thermally, so that the planet moves through this map as its core heats and cools.


\begin{figure}
\centering
\includegraphics[width=\columnwidth]{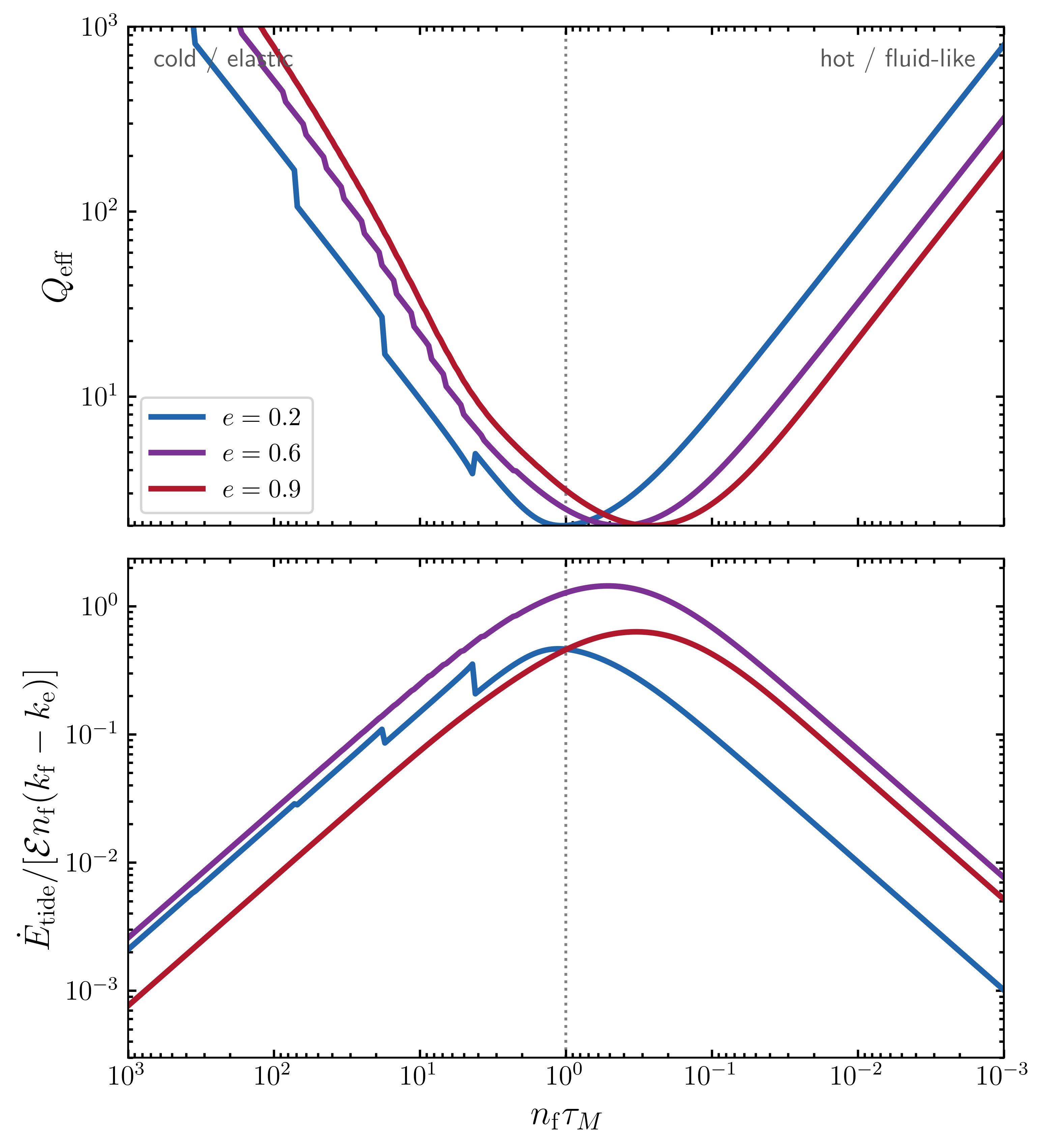}
\caption{Effective tidal dissipation and heating as a function of the tidal relaxation time for fixed eccentricities. Top: effective tidal quality factor $\qeff$. Bottom: dimensionless tidal heating rate $\dot E_{\rm tide}/[\mathcal{E}n_f(k_f-k_e)]$. The vertical dotted line marks $\nf\taum=1$. Tidal dissipation peaks near the transition between elastic and fluid-like behavior, producing a minimum in $\qeff$ and a corresponding maximum in heating.}
\label{fig:tau_sweep}
\end{figure}

\subsection{Crossing the Maxwell peak}

Figure~\ref{fig:tau_sweep} shows the complementary view to Figure~\ref{fig:q_heat}. Here the eccentricity is held fixed while the Maxwell relaxation time is varied. The horizontal axis therefore measures the location of the tidal spectrum relative to the Maxwell peak.

The upper panel shows the effective quality factor. For all eccentricities, $Q_{\rm eff}$ reaches a minimum near $n_f\tau_M\sim1$, corresponding to the transition between the elastic and fluid-like regimes. Away from this transition, the dissipation weakens and $Q_{\rm eff}$ increases approximately linearly with $n_f\tau_M$ in the elastic regime ($n_f\tau_M\gg1$) and inversely with $n_f\tau_M$ in the fluid-like regime ($n_f\tau_M\ll1$), as expected for a Maxwell body. At low eccentricity, the curves exhibit a sequence of abrupt transitions associated with changes in the highest-frequency stable zero-torque spin equilibrium. These transitions reorganize the harmonic spectrum and produce the discontinuities visible in both panels.

The lower panel shows the corresponding tidal heating rate. Dissipation is maximized when the dominant dissipation-weighted forcing frequencies lie near the Maxwell transition, $|\omega|\tau_M\sim1$. The broad maximum partly reflects the broad Maxwell response itself, but it is further broadened at high eccentricity because the tidal power is distributed over a wider range of Hansen/Kaula harmonics. As the eccentricity increases, the dominant harmonics and the equilibrium spin shift, moving the heating maximum toward smaller values of $n_f\tau_M$. Thus planets at different eccentricities encounter peak dissipation at somewhat different rheological states.

Most importantly, Figure~\ref{fig:tau_sweep} demonstrates that the tidal heating depends sensitively on the Maxwell relaxation time. Since $\tau_M$ is controlled by the temperature-dependent viscosity, tidal heating can modify its own efficiency by changing the thermal state of the core. This feedback motivates the coupled thermal evolution calculations considered in the following section.


\begin{figure*}
\centering
\includegraphics[scale=0.47]{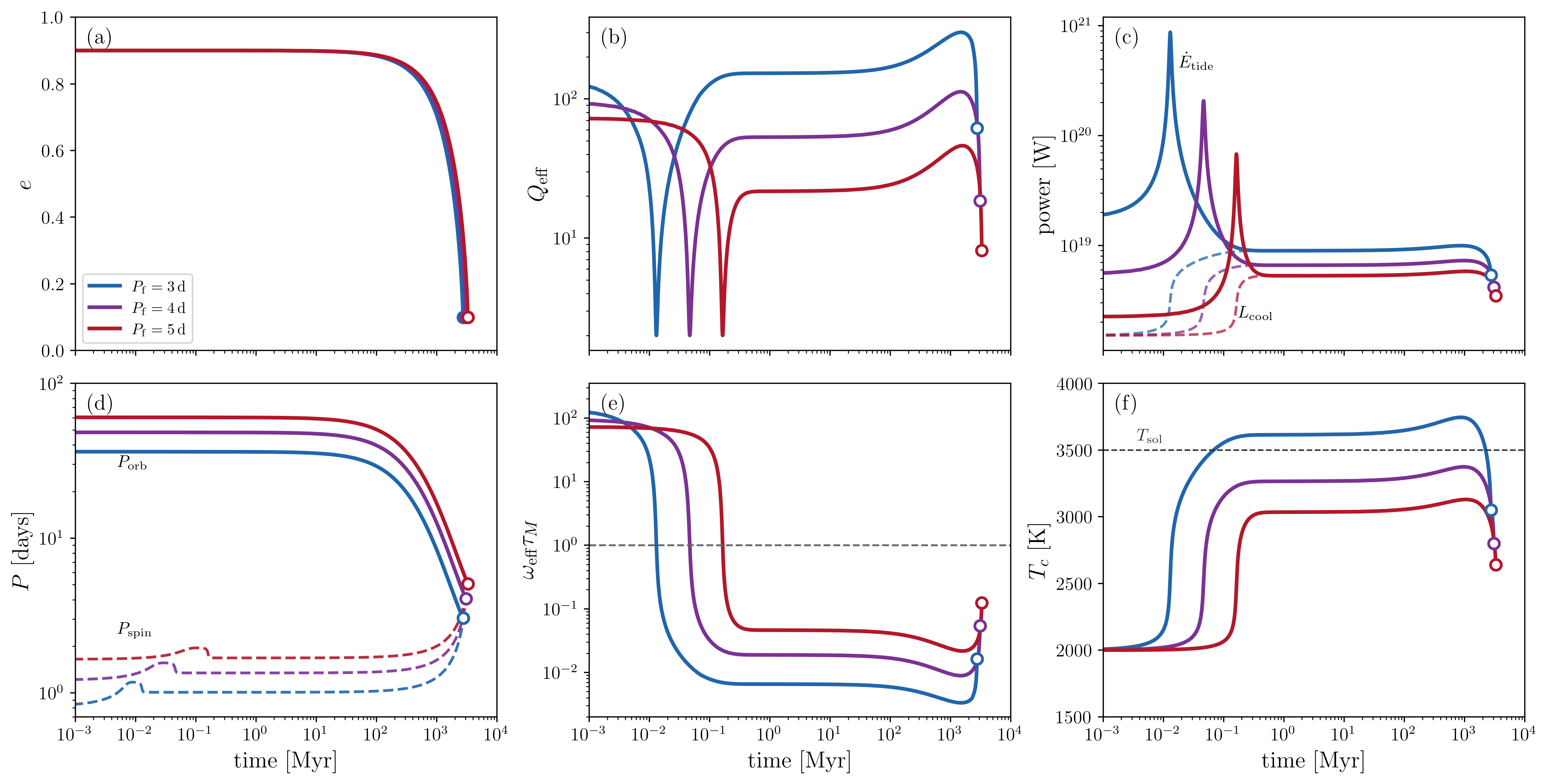}
\caption{
Thermally coupled tidal migration tracks for eccentric Neptune-like planets with final circular periods $P_f=3$, $4$, and $5\,{\rm d}$. The integrations begin at $e=0.9$ and $\tc=2000\,{\rm K}$. Panels show the evolution of (a) eccentricity, (b) effective tidal quality factor, (c) tidal heating and cooling, (d) orbital period and instantaneous equilibrium spin period, (e) $\omeff\taum$, and (f) effective core temperature. The tidal response uses $k_f=0.9$, while $k_e$, $k_f-k_e$, and $\taum$ are recomputed from the instantaneous shear modulus and gravitational rigidity of the core. Open circles mark where the integrations are terminated at $e=0.1$. Dashed curves in panel (d) show $P_{\rm spin}=2\pi/\Omega_s$, while solid curves show $P_{\rm orb}$. The horizontal dashed line in panel (f) marks the adopted solidus temperature. The tracks rapidly approach a quasi-steady state in which tidal heating balances cooling, and then circularize after a few Gyr with only very weak dependence on $P_f$.
}
\label{fig:thermal_evolution}
\end{figure*}

\section{Thermally regulated migration}
\label{sec:thermal}

We now evolve the coupled eccentricity and core temperature of a Neptune-like planet using the tidal prescription defined in Section~\ref{sec:formalism}. The purpose of this calculation is not to construct a detailed interior model, but to illustrate the feedback between tidal heating, the temperature-dependent tidal response, and orbital circularization. The fiducial parameters used in the integrations are summarized in Table~\ref{tab:thermal_params}.

\begin{table}
\centering
\caption{Fiducial parameters for the thermally coupled migration calculations in Figure~\ref{fig:thermal_evolution}.}
\label{tab:thermal_params}
\begin{tabular}{lll}
\hline
Parameter & Value & Description \\
\hline
$M_p$ & $15\,M_\oplus$ & planet mass \\
$M_c$ & $10\,M_\oplus$ & dissipating core mass \\
$R_c$ & $2\,R_\oplus$ & core radius \\
$k_f$ & $0.9$ & fluid Love number \\
$\mu_{\rm grav}$ & $2.3\times10^{11}\,{\rm Pa}$ & gravitational rigidity \\
$c_p$ & $1200\,{\rm J\,kg^{-1}\,K^{-1}}$ & specific heat \\
$\eta_{\rm ref}$ & $10^{17}\,{\rm Pa\,s}$ & viscosity at $T_{\rm ref}$ \\
$T_{\rm ref}$ & $2000\,{\rm K}$ & reference temperature \\
$E_a$ & $400\,{\rm kJ\,mol^{-1}}$ & activation energy \\
$\mu_{\rm solid}$ & $5\times10^{10}\,{\rm Pa}$ & solid shear modulus \\
$\mu_{\rm min}$ & $5\times10^8\,{\rm Pa}$ & melt-weakened shear mod. \\
$T_{\rm sol}$ & $3500\,{\rm K}$ & effective solidus temp. \\
$T_{\rm liq}$ & $6000\,{\rm K}$ & effective liquidus temp.  \\
$t_{\rm cool}$ & $3\,{\rm Myr}$ & effective cooling time \\
$\beta$ & $3$ & cooling-law index \\
$e_0$ & $0.9$ & initial eccentricity \\
$T_{c,0}$ & $2000\,{\rm K}$ & initial core temperature \\
$P_f$ & $3,4,5\,{\rm d}$ & final circular period \\
\hline
\end{tabular}
\end{table}

The orbit evolves along the constant-angular-momentum track in Equation~\eqref{eq:constant_J}. The eccentricity evolution therefore follows from orbital energy loss,
\begin{equation}
\frac{da}{dt}
=
-\frac{2a^2\dot E_{\rm tide}}{GM_\star M_p},
\qquad
\frac{de}{dt}
=
\frac{1-e^2}{2ae}\frac{da}{dt}.
\label{eq:orb_evol}
\end{equation}
The core temperature evolves according to
\begin{equation}
C_c\frac{d\tc}{dt}
=
\dot E_{\rm tide}-L_{\rm cool},
\qquad
C_c=M_c c_p .
\label{eq:thermal_evol}
\end{equation}

At each timestep, $\dot E_{\rm tide}$ is evaluated from the full harmonic sum in Equation~\eqref{eq:edot}. For computational efficiency, the Hansen amplitudes $\mathcal{F}_{mN}(e)$ are precomputed on a grid in eccentricity and interpolated during the integration. The spin is assumed to remain on the stable zero-torque branch. Rather than solving the torque equation at every timestep, we use the fitting expression for $f_{\rm eq}$ derived in Appendix~\ref{sec:app_spin}.\footnote{The fitting expression reproduces the numerical zero-torque equilibria to within $\lesssim10\%$ for $e>0.7$ over the range of $\nf\taum$ relevant here, while the limiting value $f_{\rm eq}=f_{\rm Hut}$ is recovered exactly for $\nf\taum\ll1$. The approximation is therefore used to capture the high-eccentricity transition from elastic to fluid-like response. During the later circularization phase the response lies close to the fluid-like/Hut limit and is insensitive to the detailed resonant structure of the solid-like branch.}
The forcing frequencies $\omega_{mN}$ are therefore updated self-consistently as $e$, $n$, $\tc$, and $\taum$ evolve.

The viscosity is modeled with an Arrhenius law,
\begin{equation}
\eta(\tc)
=
\eta_{\rm ref}
\exp\left[
\frac{E_a}{R_g}
\left(
\frac{1}{\tc}
-
\frac{1}{T_{\rm ref}}
\right)
\right],
\label{eq:arrhenius}
\end{equation}
with upper and lower bounds imposed for numerical stability. The shear modulus is allowed to weaken at high temperature using the phenomenological melt-fraction proxy
\begin{equation}
\phi(\tc)=
\begin{cases}
0, & \tc\leq T_{\rm sol},\\[3pt]
(\tc-T_{\rm sol})/(T_{\rm liq}-T_{\rm sol}), & T_{\rm sol}<\tc<T_{\rm liq},\\[3pt]
1, & \tc\geq T_{\rm liq},
\end{cases}
\label{eq:melt_fraction}
\end{equation}
and
\begin{equation}
\mu(\tc)
=
\mu_{\rm min}
+
\left(\mu_{\rm solid}-\mu_{\rm min}\right)(1-\phi)^2 .
\label{eq:mu_T}
\end{equation}

The tidal relaxation time and Love-number contrast are then computed from the self-gravitating Maxwell response in Section~\ref{sec:formalism},
\begin{equation}
\taum(\tc)
=
\frac{\eta(\tc)}{\mu(\tc)}
+
\frac{\eta(\tc)}{\mu_{\rm grav}},
\label{eq:tauM_T}
\end{equation}
and
\begin{equation}
k_f-k_e(\tc)
=
k_f
\frac{\mu(\tc)}
{\mu(\tc)+\mu_{\rm grav}} .
\label{eq:kcontrast_T}
\end{equation}
Thus both the location of the Maxwell peak and the amplitude of the dissipative Love-number response evolve with the thermal state of the core. For the fiducial solid shear modulus, $k_f-k_e\simeq0.16$ below the adopted solidus.

We adopt a minimal one-zone cooling law,
\begin{equation}
L_{\rm cool}
=
\frac{C_c T_{\rm ref}}{t_{\rm cool}}
\left(
\frac{\tc}{T_{\rm ref}}
\right)^\beta 
=
L_{\rm cool, 0} \left(
\frac{\tc}{T_{\rm ref}}
\right)^\beta .
\label{eq:cooling}
\end{equation}
Here $t_{\rm cool}$ and $\beta$ are effective parameters describing uncertain heat transport through the core-envelope boundary and overlying envelope. For the fiducial values listed in Table~\ref{tab:thermal_params}, $\beta=3$ and $t_{\rm cool}=3\,{\rm Myr}$, the effective core temperature remains in the range of a few thousand kelvin while the system approaches thermal balance.

Figure~\ref{fig:thermal_evolution} shows the resulting evolution. The initial thermal adjustment is rapid compared with the Gyr migration time. Tidal heating lowers the viscosity, decreases the tidal relaxation time $\taum$, and moves the response from the elastic side of the Maxwell curve toward the fluid-like regime, as seen in the rapid drop of $\omeff\taum$ in panel~(e). After this short transient, each track settles into a long-lived quasi-steady state in which the tidal heating in panel~(c) closely tracks the cooling luminosity,
\begin{equation}
\dot E_{\rm tide}\simeq L_{\rm cool}.
\end{equation}
The corresponding core temperatures in panel~(f) remain of order a few thousand kelvin. The shortest-period track briefly exceeds the adopted solidus temperature, while the wider tracks remain mostly below it. Thus the quasi-steady state does not require the core to become partially molten; partial melting acts mainly as a high-temperature regularization in the hottest cases.

Panel~(d) shows the orbital period together with the instantaneous equilibrium spin period. During the eccentric phase the planet spins much faster than the instantaneous orbital period, as expected for pseudo-synchronous rotation at high eccentricity. As the orbit circularizes, $f_{\rm eq}\rightarrow1$ and the spin period approaches the orbital period. The brief initial feature in $P_{\rm spin}$ reflects the rapid thermal change in $\taum$ at nearly fixed orbit; a finite spin-adjustment time would smooth this transient without affecting the subsequent migration.

The main result is that the three tracks reach low eccentricity on almost indistinguishable few-Gyr timescales, even though their final circular periods differ by nearly a factor of two. In a fixed-$\Delta t$ equilibrium tide, one would expect $t_e\propto a_f^8\propto P_f^{16/3}$, implying a factor of $\simeq15$ difference between $P_f=3$ and $5\,{\rm d}$. The thermally regulated tracks are much closer together because $Q_{\rm eff}$ is not fixed: stronger tidal forcing at smaller $P_f$ heats the core more efficiently, pushes the response farther into the fluid-like regime, and raises the effective tidal quality factor, as shown in panel~(b). In the next section we show that this compensation can reduce the scaling to a much weaker dependence, of order $t_e\propto P_f^{0.3}$ for the fiducial quasi-steady parameters (see also \ref{secc:app_quasi_steady}).


\begin{figure*}
\centering
\includegraphics[scale=0.6]{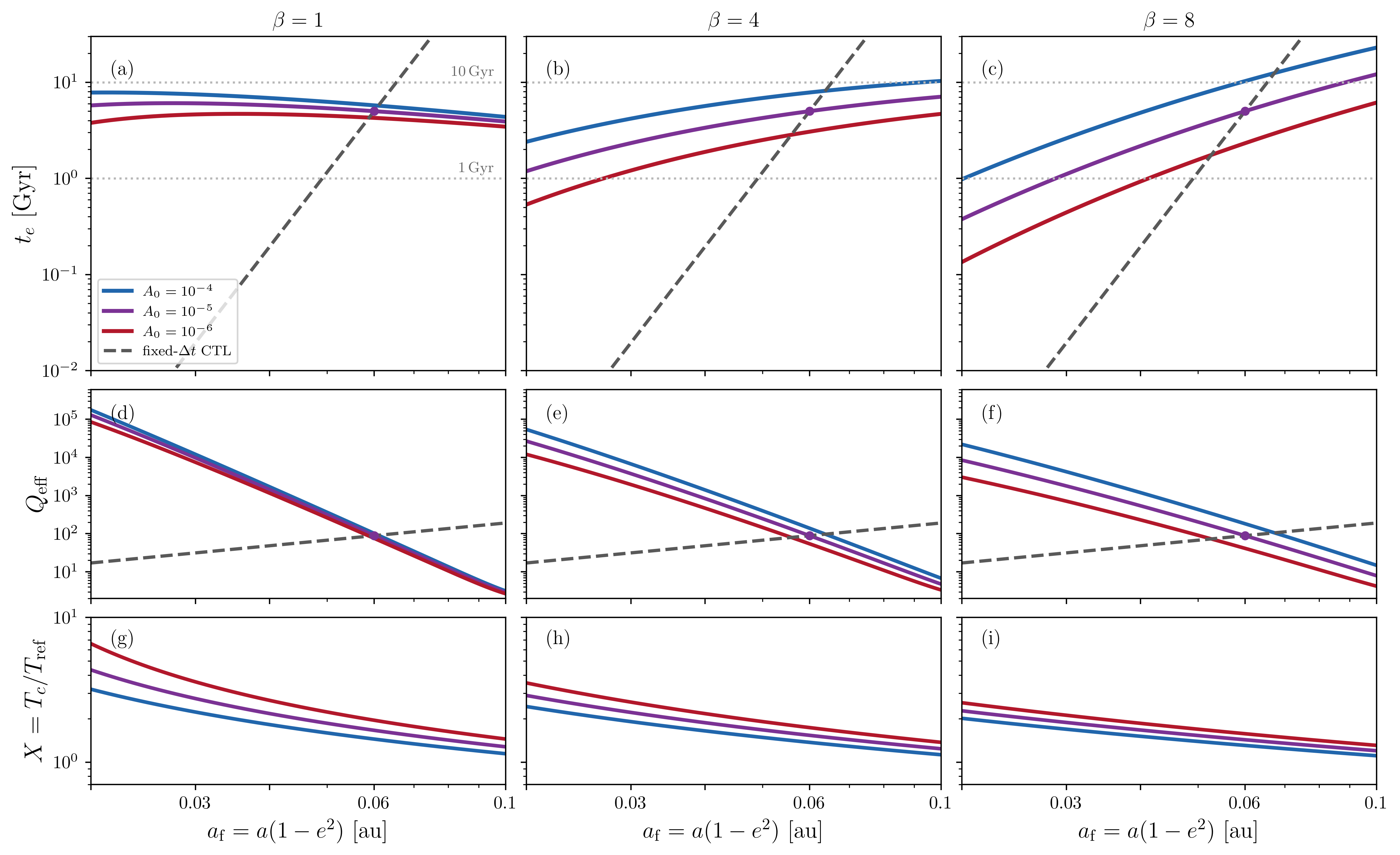}
\caption{
Quasi-steady thermal regulation along a constant-angular-momentum migration track at fixed eccentricity, $e=0.7$. Columns correspond to cooling indices $\beta=1$, 4, and 8 in Equation~\eqref{eq:cooling}. Colored curves show solutions of the thermal-balance equation for three values of the dimensionless normalization $A_0$. In each column, the cooling normalization $L_{\rm cool,0}$ is chosen so that the middle model, $A_0=10^{-5}$, has $t_e=5\,{\rm Gyr}$ at $a_f=0.06\,{\rm au}$; the dashed gray fixed-$\Delta t$ comparison is normalized at the same point. Upper panels show the eccentricity damping time $t_e$, middle panels show the representative tidal quality factor $Q_{\rm eff}$, and lower panels show the corresponding equilibrium temperature $X=T_c/T_{\rm ref}$. Thermal regulation weakens the dependence of $t_e$ on $a_f$ relative to the standard constant-time-lag scaling, $t_e\propto a_f^8$, by driving larger $Q_{\rm eff}$ at smaller circularization radii.
}
\label{fig:quasi_ss}
\end{figure*}

\subsection{Quasi-steady thermal regulation}

The integrations in Figure~\ref{fig:thermal_evolution} rapidly approach a moving thermal equilibrium in which
$\dot E_{\rm tide}\simeq L_{\rm cool}$.
This state is ``quasi-steady'' because the equilibrium temperature changes slowly as the orbit circularizes. The thermal adjustment time is short compared with the eccentricity damping time, so the core temperature approximately tracks the instantaneous balance set by the current eccentricity and circularization radius.

The origin of the weak $P_f$ dependence can be seen analytically in the fluid-like Maxwell regime, $\omega\tau_M\ll1$. In this limit the tidal response reduces to a weak-friction equilibrium tide with an effective lag proportional to the gravity-controlled viscous time, $\tau_v=\eta/\mu_{\rm grav}$ (Eq. [\ref{eq:edot_hut_af_selfgrav}]). Combining this limit with the Arrhenius viscosity law and the cooling prescription in Equation~\eqref{eq:cooling} gives an implicit equation for the equilibrium temperature $T_c(a_f,e)$, derived in Appendix~\ref{secc:app_quasi_steady}. Stronger tidal forcing at smaller $a_f$ raises the equilibrium temperature; this lowers the viscosity, shortens the tidal relaxation time, and reduces the dissipation efficiency. The thermal response therefore partially compensates for the geometric strengthening of the tides.

In the quasi-steady state, the orbital energy loss is set by the cooling luminosity. From Equation~\eqref{eq:orb_evol},
\begin{equation}
t_a^{-1}
\equiv
\left|
\frac{\dot a}{a}
\right|
=
\frac{2aL_{\rm cool}}
     {GM_\star M_p}.
\end{equation}
Using the constant-angular-momentum relation, $a(1-e^2)=a_f$, together with the cooling law in Equation~\eqref{eq:cooling}, the eccentricity damping time becomes
\begin{equation}
t_e
\equiv
\frac{e}{|\dot e|}
=
\frac{
e^2GM_\star M_p
}{
a_fL_{\rm cool,0}X^\beta
},
\qquad
X\equiv T_c/T_{\rm ref}.
\end{equation}

The dependence on $a_f$ is therefore controlled by how the equilibrium temperature changes with orbital distance. Differentiating the quasi-steady balance condition in Equation~\eqref{eq:app_dimensionless} at fixed eccentricity gives
\begin{equation}
t_e \propto a_f^{s},
\qquad
s=
-1+\frac{9\beta}{\Theta/X+\beta},
\label{eq:te_scaling}
\end{equation}
where $\Theta=E_a/(R_gT_{\rm ref})$ measures the temperature sensitivity of the viscosity. For rocky and icy materials, the Arrhenius sensitivity is large, $\Theta\gg1$; for the fiducial parameters used here, $\Theta\simeq24$. Since the equilibrium temperatures are $X\simeq1$--$2$, the term $\Theta/X$ strongly suppresses the exponent $s$, making it much smaller than the constant-time-lag value $s=8$ for a broad range of cooling laws.


Figure~\ref{fig:quasi_ss} illustrates these quasi-steady solutions at fixed eccentricity, $e=0.7$, chosen as a representative point near the broad peak of tidal heating. The precise choice of eccentricity affects the normalization, but not the basic flattening mechanism. The lower panels show the equilibrium temperature for several cooling indices $\beta$ and normalizations $A_0$. The temperature increases toward smaller $a_f$, where tidal forcing is stronger. This heating lowers the tidal relaxation time and moves the response farther into the fluid-like regime, producing the larger $Q_{\rm eff}$ values shown in the middle panels.

The upper panels show the corresponding eccentricity damping time. In all cases the dependence on $a_f$ is much flatter than the constant-time-lag prediction, shown by the dashed curves. The flattening occurs because the increase in $Q_{\rm eff}$ at smaller $a_f$ partially compensates for the stronger tidal forcing. Weakly temperature-dependent cooling can make $t_e$ nearly independent of $a_f$, while steeper cooling laws yield more moderate temperatures and still substantially reduce the scaling relative to $t_e\propto a_f^8$.

\section{Implications for eccentric warm Neptunes}

Thermally regulated tides predict a much weaker dependence of circularization time on final circularization distance, or equivalently on $P_f$, than fixed-efficiency tidal models. This may help explain why eccentric short-period Neptune-mass planets are observed over a broad range of orbital distances \citep[e.g.,][]{Correia2020}: a single constant-$Q$ or constant-$\Delta t$ prescription would either circularize the innermost systems too efficiently or leave the outer systems insufficiently damped. As a preliminary check, Appendix~\ref{sec:appendix_observations} shows that the eccentric fraction of isolated hot Neptunes varies more weakly with $a_f$ than that of hot Jupiters; in the innermost bin, $a_f<0.04\,{\rm AU}$, 5 of 10 hot Neptunes have $e>0.1$, compared with only 4 of 48 hot Jupiters.

The model makes several testable predictions:
(i) eccentric hot Neptunes should persist to smaller circularization radii than expected from constant-$Q$ or constant-time-lag models;
(ii) eccentricity should correlate with signatures of recent or ongoing tidal heating, including inflated radii and enhanced atmospheric escape; and
(iii) actively migrating systems should spend extended time at high eccentricity, increasing the expected number of Neptune- and sub-Saturn-mass planets with $e\sim0.8$--$0.9$ and small pericenter distances.
GJ~3470\,b provides a suggestive example of the second prediction: it is a polar warm Neptune with $e\simeq0.12$, an evaporating atmosphere, and a radius that may be inflated by tidal heating \citep{Gummi2022}. Kepler-1656\,b \citep[$e\simeq0.84$;][]{Rubenzahl2024} may represent the more extreme high-eccentricity migration phase.

This behavior should be most relevant when dissipation is controlled by a dense viscoelastic interior rather than by an extended gaseous envelope. Although the mechanism is not limited to Neptune-like planets, the presence of an envelope may be important because it can slow cooling and help maintain the thermally regulated state. A similar feedback could operate in rocky planets and super-Earths with viscoelastic interiors, but the outcome is likely more sensitive to uncertain cooling physics, composition, melting, and volatile content. In the absence of a substantial envelope, such planets may cool more efficiently and circularize more rapidly, making long-lived eccentric phases less prevalent. Determining where the transition occurs between rapidly cooling rocky planets and thermally buffered Neptune-like planets will require coupled interior-orbital models beyond the scope of this work.

\section{Discussion}

We have presented a thermally regulated model for the tidal migration of eccentric planets whose dissipation is controlled by a temperature-dependent Maxwell core. Tidal heating raises the core temperature, lowers the viscosity, and changes the Maxwell relaxation time. The planet therefore does not migrate with a fixed dissipation efficiency: it passes through the Maxwell dissipation peak and then evolves toward a more weakly dissipative, fluid-like response.

This feedback weakens the usual dependence of circularization time on orbital distance. In fixed-$Q$ or constant-time-lag models, the damping time increases steeply with final circularization radius. In our model, stronger tidal forcing at smaller $P_f$ heats the core more efficiently and pushes it farther from the Maxwell peak, raising $Q_{\rm eff}$ and slowing the subsequent damping. This self-adjustment partially compensates the geometric scaling of the tides, allowing the tracks in Figure~\ref{fig:thermal_evolution} to circularize on comparable, few-Gyr timescales despite their different $P_f$.

Thermal--orbital feedback has been studied previously for viscoelastic planets and satellites \citep[e.g.,][]{Driscoll2015,Renaud2018,Herath2026}. Our calculation differs in focusing on the high-eccentricity migration regime, where the tidal response is spread across a broad harmonic spectrum. In this regime, the characteristic forcing frequency can differ substantially from the mean motion: Figure~\ref{fig:q_heat}(a) shows that $\omeff$ decreases as the orbit circularizes, even though $n$ increases. For highly eccentric planets, $\omeff$ can exceed $n$ by more than an order of magnitude, so replacing the spectrum by a single frequency $\sim n$ can place the planet on the wrong side of the Maxwell response.

The same point applies to the eccentricity dependence of the heating. Low-eccentricity expressions proportional to $e^2$ reproduce the behavior only near circular orbits. Along a constant-angular-momentum track, the Hut weak-friction expression peaks near $e\simeq0.74$ and then declines toward higher eccentricity because the semimajor axis increases. The frequency-dependent Maxwell response further shifts this peak depending on rheology. Thus, during high-eccentricity migration, both the heating rate and the relevant forcing frequency must be computed from the full eccentric tidal spectrum.

Several ingredients remain uncertain. Our cooling law is intentionally one-zone and does not resolve heat transport through the core and gaseous envelope. The envelope may regulate the cooling luminosity, shift the equilibrium temperature, introduce additional thermal timescales, and respond to tidal heating through radius inflation or mass loss \citep[e.g.,][]{hallat2026a,hallat2026b}. Coupling frequency-dependent viscoelastic dissipation to multi-zone structure models will therefore be needed to connect tidal migration, radius inflation, atmospheric escape, and interior evolution. More broadly, thermally regulated tides imply that present-day eccentricities, periods, radii, and ages may constrain otherwise inaccessible properties such as viscosity, melting state, and cooling efficiency.

\section{Conclusions}

We have shown that tidal migration of eccentric planets can be thermally self-regulated when dissipation is controlled by a temperature-dependent viscoelastic core. Tidal heating drives the core toward the Maxwell dissipation peak, accelerating migration, but further heating lowers the viscosity, pushes the response into the fluid-like regime, and weakens the dissipation. The planet then settles into a quasi-steady state in which the migration rate is regulated by how efficiently the interior can cool, producing long-lived eccentric phases.

A central consequence is that the eccentricity damping timescale depends only weakly on the final circularization radius compared to standard constant time-lag or constant-$Q$ models. This provides a natural explanation for why hot Neptunes may retain measurable eccentricities over a broad range of short-period orbits, while hot Jupiters show a strong decline in eccentric fraction toward small orbital separations. The model also predicts a prolonged population of actively migrating Neptune-mass planets with high eccentricities and small pericenter distances.

More broadly, thermally regulated tidal migration directly couples a planet's orbital evolution to the thermal and rheological state of its interior. Present-day eccentricities, orbital periods, radii, and system ages may therefore provide indirect constraints on otherwise inaccessible interior properties such as viscosity, melting state, and cooling efficiency.

\bigskip

\section*{acknowledgments}
The authors are grateful for helpful discussions with Douglas Lin, Gongjie Li, Yanqin Wu, Janosz Dewberry, Songhu Wang, and Diego Mu\~noz.
This work was supported by the National Science Foundation under collaborative grant AST-2511257.

\newpage

\appendix

\begin{figure*}
\centering
\begin{minipage}{0.47\textwidth}
\centering
\includegraphics[width=\linewidth]{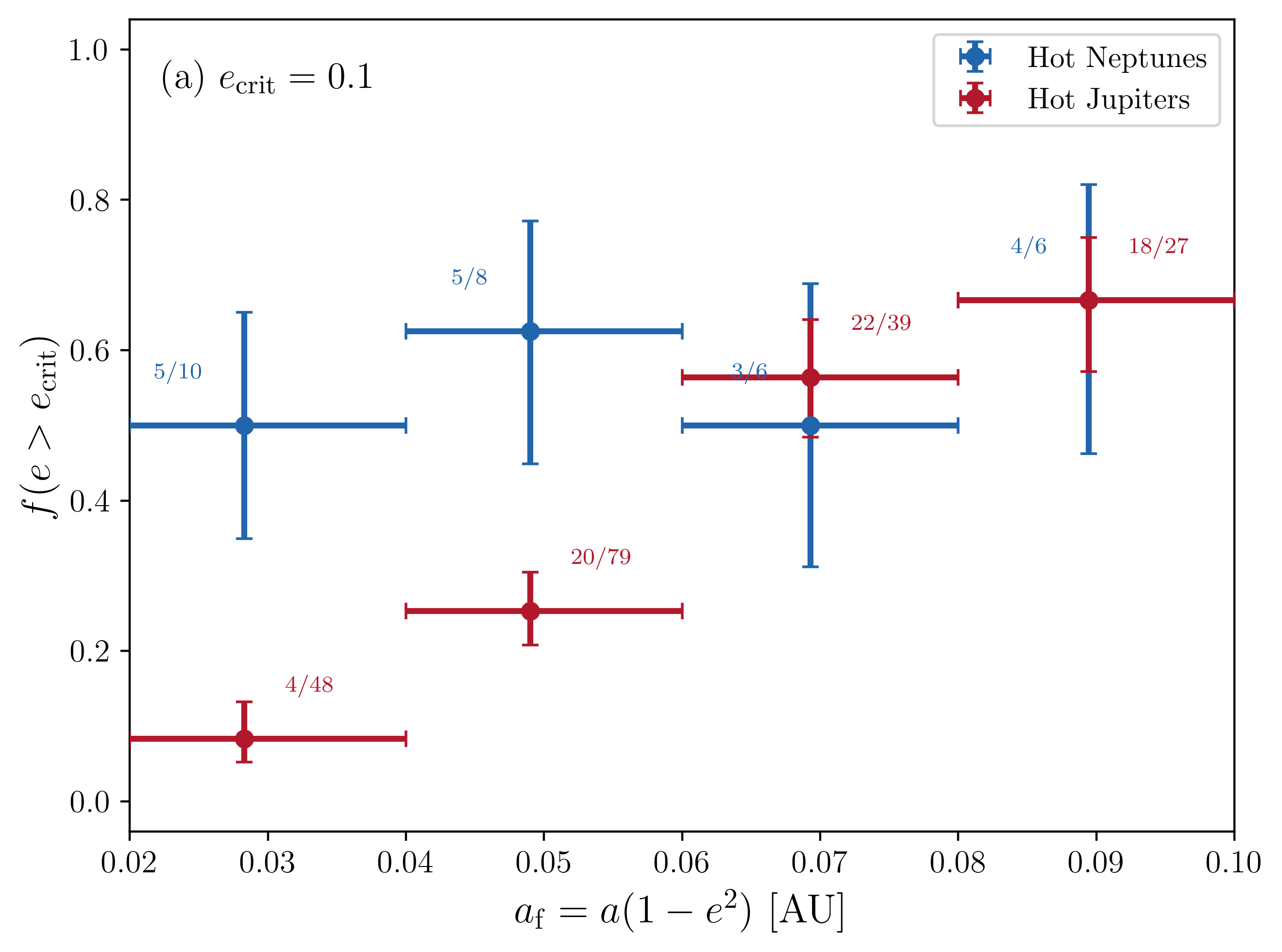}
\end{minipage}
\hfill
\begin{minipage}{0.47\textwidth}
\centering
\includegraphics[width=\linewidth]{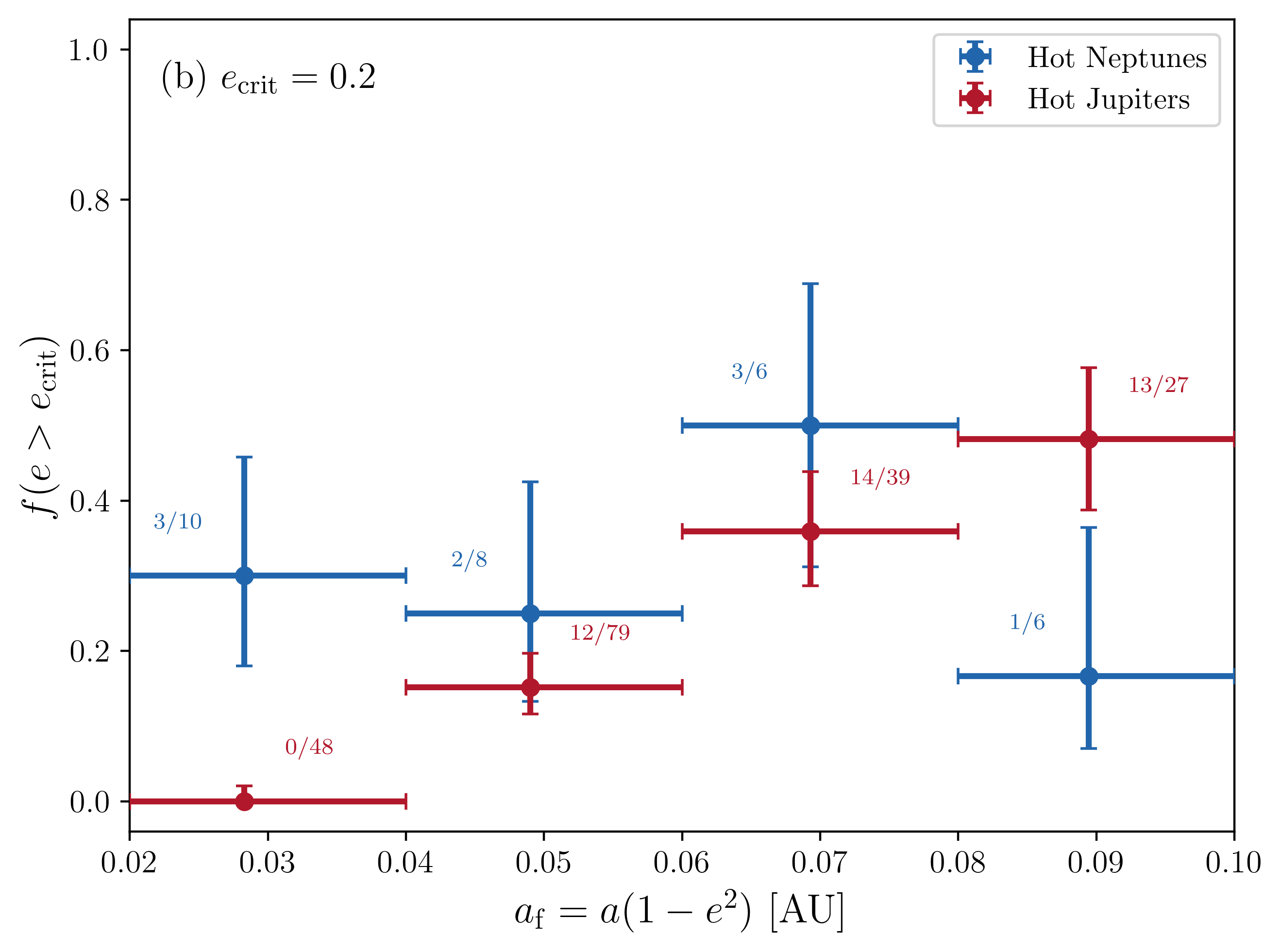}
\end{minipage}
\caption{
Fraction of isolated hot Neptunes ($10<M_p/M_\oplus<30$; blue) and hot Jupiters ($M_p>100\,M_\oplus$; red) with eccentricities exceeding a threshold $e_{\rm crit}$ as a function of inferred circularization radius, $a_f=a(1-e^2)$. The two panels correspond to $e_{\rm crit}=0.1$ and $0.2$. Points show fixed-width $0.02\,{\rm AU}$ bins, and labels give the number of eccentric planets relative to the total number of systems in each bin. Systems are required to have reported eccentricity uncertainties $\sigma_e<0.1$, and hosts with known planetary companions inside $0.5\,{\rm AU}$ are excluded. Hot Jupiters show a pronounced decline in eccentric fraction toward smaller $a_f$, while hot Neptunes show little evidence for a comparable trend. Although the current hot-Neptune sample remains small, this behavior is qualitatively consistent with the weak dependence of circularization time on final orbital distance predicted by thermally regulated tidal evolution.
}
\label{fig:ecc_fraction_obs}
\end{figure*}

\section{Observed eccentric fractions of hot Neptunes and hot Jupiters}
\label{sec:appendix_observations}

The thermally regulated migration model developed in this work predicts a weaker dependence of circularization time on final orbital distance than standard equilibrium-tide models. As a preliminary observational check, we compare the eccentric fractions of hot Neptunes and hot Jupiters as a function of the inferred circularization radius, $a_f=a(1-e^2)$.

We queried the NASA Exoplanet Archive Composite Parameters table (\texttt{pscomppars}) on 2026 June 3. For each planet, we retrieved the planet name, host name, mass, orbital period, semimajor axis, eccentricity, eccentricity uncertainties, stellar mass, and stellar age. We selected planets with $P<30\,{\rm d}$, measured semimajor axes, and reported eccentricities. We define hot Neptunes as $10<M_p/M_\oplus<30$ and hot Jupiters as $M_p>100\,M_\oplus$. We further require $0.02<a_f/{\rm AU}<0.10$, $0.02<a/{\rm AU}<0.30$, and $e<0.95$.

To focus on isolated systems that may have experienced high-eccentricity migration, we remove planets with known planetary companions inside $0.5\,{\rm AU}$ around the same host. Companion semimajor axes are taken from the archive when available and otherwise estimated from the orbital period and stellar mass. We also require reported eccentricity uncertainties with $\sigma_e<0.1$, where $\sigma_e$ is the larger of the upper and lower reported errors. This removes many systems for which circular orbits were assumed or eccentricities were poorly documented.

Figure~\ref{fig:ecc_fraction_obs} shows the fraction of systems with eccentricities above $e_{\rm crit}=0.1$ and $0.2$, using fixed-width $0.02\,{\rm AU}$ bins in $a_f$. The hot-Jupiter sample shows a pronounced decline in eccentric fraction toward smaller $a_f$, consistent with efficient circularization at small orbital distances. In contrast, the eccentric fraction of hot Neptunes shows little evidence for a comparable decline. Although the hot-Neptune sample remains small, this behavior is qualitatively consistent with the weaker $a_f$ dependence predicted by thermally regulated tidal evolution. We also examined higher eccentricity thresholds, but the current samples become small enough that the resulting trends are dominated by counting noise.

The persistence of eccentric hot Neptunes at small circularization radii is driven by a few well-characterized systems. In the innermost bin, $a_f<0.04\,{\rm AU}$, the selected eccentric hot Neptunes for $e_{\rm crit}=0.1$ include K2-25 b, GJ~436 b, GJ~724 b, HD~134060 b, and GJ~674 b, with eccentricities ranging from $e\simeq0.14$ to $e\simeq0.58$. Several of these eccentricities are measured with high significance; GJ~436 b remains the archetypal close-in Neptune with a substantial eccentricity.

As a non-parametric check, we compared the $a_f$ distributions of eccentric and non-eccentric systems within each mass class for $e_{\rm crit}=0.1$ using two-sample KS tests. For hot Neptunes, the two distributions are statistically indistinguishable ($D=0.18$, $p=0.92$). For hot Jupiters, eccentric systems are strongly shifted toward larger $a_f$ ($D=0.46$, $p=9.2\times10^{-9}$). Thus the contrast seen in Figure~\ref{fig:ecc_fraction_obs} does not appear to be an artifact of the adopted binning.

Given the limited size of the current sample of hot Neptunes and the heterogeneous nature of eccentricity measurements, Figure~\ref{fig:ecc_fraction_obs} should be viewed as a qualitative consistency check rather than a definitive test. Future radial-velocity surveys and transit follow-up observations will enlarge the sample of eccentric hot Neptunes and enable a more rigorous comparison with thermally regulated tidal evolution.


\begin{figure}
    \centering
    \includegraphics[width=0.78\linewidth]{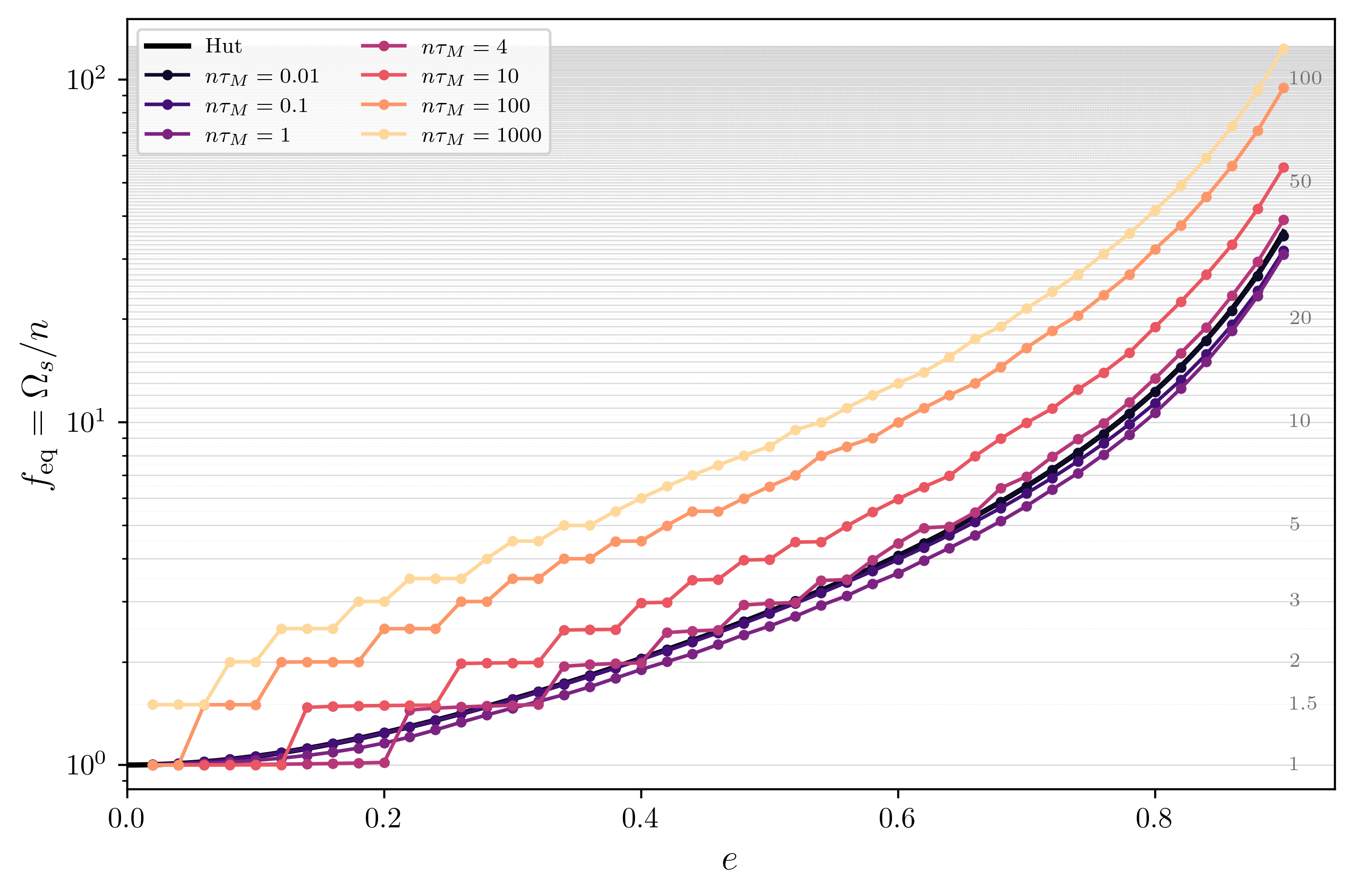}
    \caption{
    Equilibrium spin frequency \(f_{\rm eq}=\Omega_s/n\) as a function of
    eccentricity for fixed values of the local Maxwell parameter \(n\taum\).
    The black curve shows the Hut pseudo-synchronous value. Horizontal guide
    lines mark the approximate spin-orbit resonances
    \(\Omega_s/n=N/2\) associated with the dominant \(m=2\) harmonics. In the
    solid-like regime, the equilibrium spin can lock near these resonances.
    At lower eccentricity the resonant structure is visibly discrete, whereas
    at high eccentricity the relevant \(N\) values are large and densely spaced,
    producing a smoother envelope.
    }
    \label{fig:spin_equilibria_resonances}
\end{figure}

\begin{figure}
    \centering
    \includegraphics[width=0.6\linewidth]{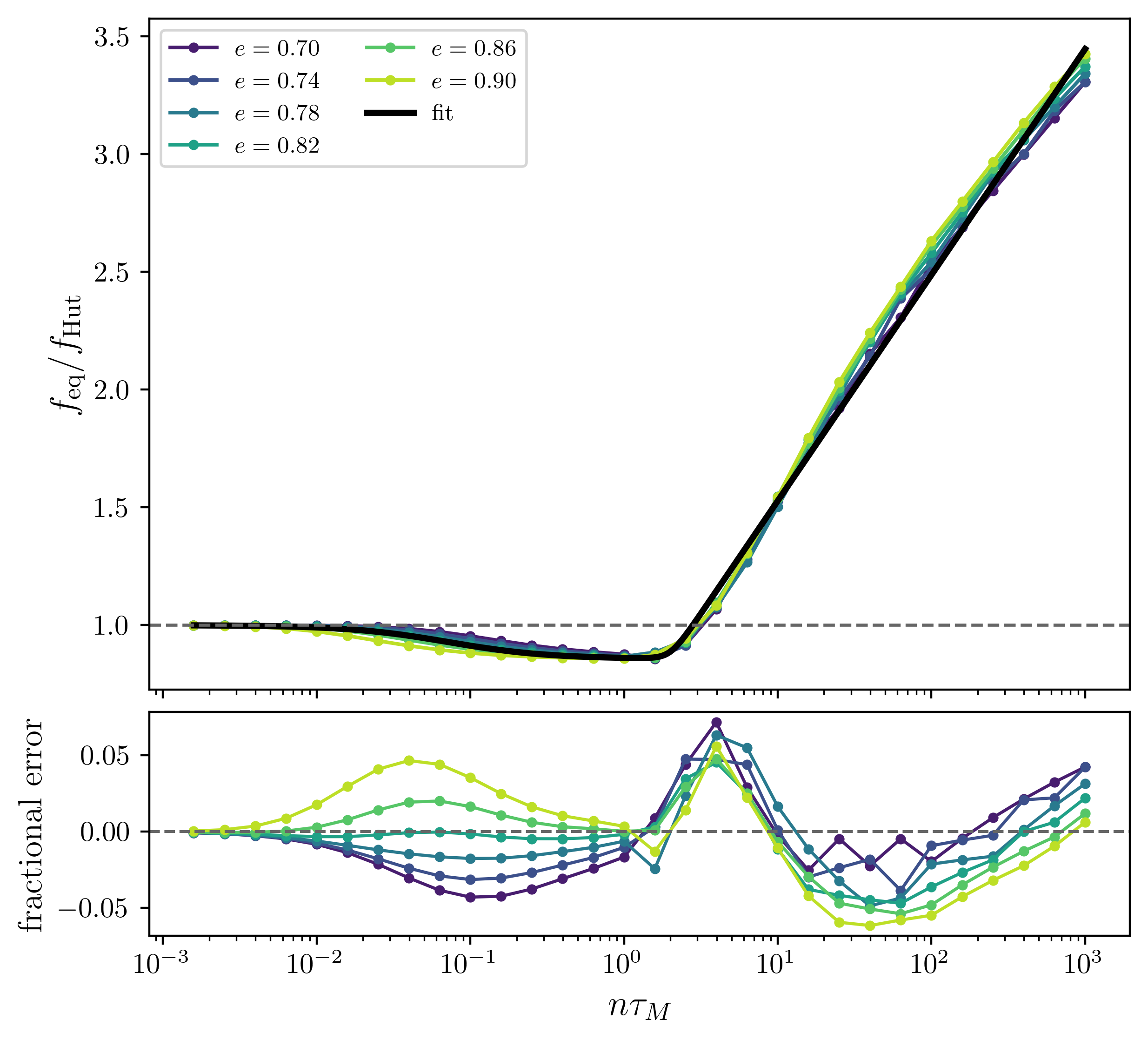}
    \caption{
    High-eccentricity collapse of the equilibrium spin, normalized by the Hut
    value, as a function of \(x=n\taum\). Points show numerical zero-torque
    equilibria for \(0.7\le e\le0.9\), selecting the highest-frequency stable
    branch when multiple stable roots exist. The black curve shows the
    empirical envelope fit in Eq.~\eqref{eq:spin_equilibrium_fit}. The lower
    panel shows the fractional residuals. The fit is intended for the early
    high-eccentricity transition from solid-like to fluid-like response.
    }
    \label{fig:spin_equilibrium_fit}
\end{figure}

\section{Spin equilibria}
\label{sec:app_spin}

In the main text we assume that the planetary spin rapidly relaxes to a
stable zero-torque equilibrium,
\begin{equation}
    \Omega_s = f_{\rm eq}(e,\nf\taum)\,n .
\end{equation}
Here we describe how \(f_{\rm eq}\) is computed.

The tidal torque follows from the same harmonic expansion used for the heating
rate in Eq.~\eqref{eq:edot}. Up to an overall positive normalization, the
spin torque can be written as
\begin{equation}
    \mathcal{T}(\Omega_s)
    =
    \sum_{m,N}
    m\,\mathcal{F}_{mN}(e)\,
    \mathrm{Im}\!\left[k_2(\omega_{mN})\right],
    \label{eq:spin_torque}
\end{equation}
where \(\mathcal{F}_{mN}\) and \(\omega_{mN}=Nn-m\Omega_s\) are defined in
Eqs.~\eqref{eq:fmn} and \eqref{eq:omega_mn}. The \(m=0\) terms do not
contribute to the spin torque because of the prefactor \(m\), so the torque is
set by the \(m=\pm2\) harmonics. We include both positive and negative \(N\)
in the torque sum.

For each pair \((e,\nf\taum)\), we find all roots of
\begin{equation}
    \mathcal{T}(\Omega_s)=0,
\end{equation}
and retain the stable ones satisfying
\begin{equation}
    \left.
    \frac{\partial \mathcal{T}}{\partial \Omega_s}
    \right|_{\Omega_s=\Omega_{\rm eq}}
    <0 .
\end{equation}
When multiple stable roots are present, we select the highest-frequency stable
branch. This prescription recovers the Hut pseudo-synchronous spin in the
fluid-like/weak-friction limit, \(\nf\taum\ll1\).

Using the constant-angular-momentum relation in Eq.~\eqref{eq:constant_J},
the local Maxwell parameter is
\begin{equation}
    x \equiv n\taum
    =
    \left(1-e^2\right)^{3/2}\nf\taum .
    \label{eq:x_ntau}
\end{equation}
Although \(\nf\taum\) is the natural parameter for a fixed final circular orbit,
the local response of the Maxwell body is controlled by \(\omega_{mN}\taum\),
and therefore by \(n\taum\) at fixed spin ratio \(\Omega_s/n\).

Figure~\ref{fig:spin_equilibria_resonances} shows \(f_{\rm eq}=\Omega_s/n\)
as a function of eccentricity for several fixed values of \(n\taum\). In the
solid-like regime, the torque is organized by spin-orbit resonances. For the
dominant \(m=2\) harmonics these occur approximately when
\begin{equation}
    \omega_{2N}=Nn-2\Omega_s\simeq0,
    \qquad
    \frac{\Omega_s}{n}\simeq\frac{N}{2}.
\end{equation}
At modest eccentricities these resonances appear as discrete plateaus or jumps.
At high eccentricity, however, the relevant Hansen power is spread over many
large \(N\), so the resonances become densely spaced and the envelope appears
nearly continuous.

Because the solid-like branch is intrinsically resonant, we do not attempt to
fit the entire eccentricity range with a smooth function. For the early
transition from solid-like to fluid-like response, however, the orbit remains
highly eccentric. In the range \(0.7\lesssim e\lesssim0.9\), the numerical
equilibria are well represented by a simple envelope fit in terms of
\(x=n\taum\):
\begin{equation}
    \frac{f_{\rm eq}}{f_{\rm Hut}}
    \simeq
    1
    - \frac{1}{7\left[1+\left(14x\right)^{-4/3}\right]}
    + \frac{1}{24}
      \ln\left[
        1+\left(\frac{x}{2}\right)^{10}
      \right].
    \label{eq:spin_equilibrium_fit}
\end{equation}
This expression should be interpreted as a compact approximation to the
high-eccentricity envelope, not as a model for individual spin-orbit resonances.


\section{Quasi-steady thermal balance}
\label{secc:app_quasi_steady}

The numerical integrations in Section~\ref{sec:thermal} rapidly approach a state in which tidal heating approximately balances cooling,
\begin{equation}
\dot E_{\rm tide}\simeq L_{\rm cool},
\end{equation}
while the eccentricity evolves on a much longer timescale. This motivates a quasi-steady estimate in which the core temperature adjusts to the slowly evolving tidal forcing.

In the fluid-like branch of the Maxwell response, $|\omega|\taum\ll1$, Equation~\eqref{eq:maxwell} gives
\begin{equation}
\mathrm{Im}\!\left[k_2(\omega)\right]
\simeq
(k_f-k_e)\omega\taum .
\end{equation}
Using Equation~\eqref{eq:kcontrast},
\begin{equation}
(k_f-k_e)\taum
=
k_f\frac{\eta}{\mu_{\rm grav}} .
\end{equation}
Thus, in the weak-friction limit, the tidal response is controlled by the gravity-controlled viscous relaxation time.

For a zero-torque equilibrium spin state, the weak-friction heating rate is
\begin{equation}
\dot E_{\rm tide}
=
3k_f
\frac{\eta}{\mu_{\rm grav}}
\frac{G^2M_\star^3R_c^5}{a^9}
\frac{F_{\rm Hut}(e)}{(1-e^2)^{15/2}} .
\label{eq:edot_hut_selfgrav}
\end{equation}
Along a constant-angular-momentum track, $a(1-e^2)=a_f$, this becomes
\begin{equation}
\dot E_{\rm tide}
=
3k_f
\frac{\eta}{\mu_{\rm grav}}
\frac{G^2M_\star^3R_c^5}{a_f^9}
(1-e^2)^{3/2}F_{\rm Hut}(e).
\label{eq:edot_hut_af_selfgrav}
\end{equation}

We define
\begin{equation}
X\equiv \frac{T_c}{T_{\rm ref}},
\qquad
\Theta\equiv \frac{E_a}{R_gT_{\rm ref}}
\simeq
24
\left(\frac{E_a}{400\,{\rm kJ\,mol^{-1}}}\right)
\left(\frac{T_{\rm ref}}{2000\,{\rm K}}\right)^{-1},
\label{eq:X_Theta_defs}
\end{equation}
so that the Arrhenius viscosity becomes
\begin{equation}
\eta(T_c)
=
\eta_{\rm ref}
\exp\left[
\Theta\left(\frac{1}{X}-1\right)
\right].
\label{eq:eta_X}
\end{equation}

Substituting Equation~\eqref{eq:eta_X} into the thermal-balance condition and using $L_{\rm cool}=L_{\rm cool,0}X^\beta$ gives
\begin{equation}
\frac{
\exp\left[\Theta(X^{-1}-1)\right]
}
{X^\beta}
=
A_0
\left(\frac{a_f}{a_{f,0}}\right)^9
\frac{1}{(1-e^2)^{3/2}F_{\rm Hut}(e)}.
\label{eq:app_dimensionless}
\end{equation}
The dimensionless normalization is
\begin{align}
A_0
&\equiv
\frac{
L_{{\rm cool},0}\mu_{\rm grav}a_{f,0}^9
}
{
3k_f\eta_{\rm ref}G^2M_\star^3R_c^5
}
\nonumber\\
&\simeq
4.5\times10^{-6}
\left(\frac{L_{{\rm cool},0}}{10^{18}\,{\rm W}}\right)
\left(\frac{\mu_{\rm grav}}{2\times10^{11}\,{\rm Pa}}\right)
\left(\frac{k_f}{0.9}\right)^{-1}
\left(\frac{\eta_{\rm ref}}{10^{17}\,{\rm Pa\,s}}\right)^{-1}
\left(\frac{M_\star}{M_\odot}\right)^{-3}
\left(\frac{R_c}{2R_\oplus}\right)^{-5}
\left(\frac{a_{f,0}}{0.05\,{\rm au}}\right)^9 .
\label{eq:app_A0}
\end{align}

The eccentricity damping time is
\begin{equation}
t_e
\equiv
\frac{e}{|\dot e|}
=
\frac{e^2GM_\star M_p}{a_f\dot E_{\rm tide}} .
\label{eq:te_def_app}
\end{equation}
Using thermal balance, this becomes
\begin{equation}
t_e
=
\frac{
e^2GM_\star M_p
}{
a_fL_{\rm cool,0}X^\beta
}.
\label{eq:te_qss}
\end{equation}
The dependence of $t_e$ on $a_f$ is therefore set by how the equilibrium temperature varies with orbital distance. Differentiating Equation~\eqref{eq:app_dimensionless} at fixed eccentricity gives the local power-law index
\begin{equation}
\frac{d\ln t_e}{d\ln a_f}
=
-1+
\frac{9\beta}{\beta+\Theta/X}.
\label{eq:te_powerlaw_index}
\end{equation}
For the fiducial parameters, $\Theta\simeq24$. With $\beta=3$ and equilibrium temperatures $X\simeq1.5$--$1.8$, Equation~\eqref{eq:te_powerlaw_index} gives
\begin{equation}
t_e\propto a_f^{0.4-0.7},
\qquad
t_e\propto P_f^{0.3-0.5}.
\end{equation}
This explains the weak period dependence found in Figure~\ref{fig:thermal_evolution}, and is much flatter than the fixed-$\Delta t$ scaling,
$t_e\propto a_f^8\propto P_f^{16/3}$.

\bibliography{refs}

\begin{thebibliography}{}
\expandafter\ifx\csname natexlab\endcsname\relax\def\natexlab#1{#1}\fi
\providecommand{\url}[1]{\href{#1}{#1}}
\providecommand{\dodoi}[1]{doi:~\href{http://doi.org/#1}{\nolinkurl{#1}}}
\providecommand{\doeprint}[1]{\href{http://ascl.net/#1}{\nolinkurl{http://ascl.net/#1}}}
\providecommand{\doarXiv}[1]{\href{https://arxiv.org/abs/#1}{\nolinkurl{https://arxiv.org/abs/#1}}}

\bibitem[{{Albrecht} {et~al.}(2022){Albrecht}, {Dawson}, \& {Winn}}]{Albrecht2022}
{Albrecht}, S.~H., {Dawson}, R.~I., \& {Winn}, J.~N. 2022, \pasp, 134, 082001, \dodoi{10.1088/1538-3873/ac6c09}

\bibitem[{{Attia} {et~al.}(2023){Attia}, {Petigura}, \& {Crossfield}}]{Attia2023}
{Attia}, O., {Petigura}, E.~A., \& {Crossfield}, I. J.~M. 2023, \aj, 166, 240, \dodoi{10.3847/1538-3881/acf5ea}

\bibitem[{{Bourrier} {et~al.}(2018){Bourrier}, {Lovis}, {Beust}, {Ehrenreich}, {Henry}, {Astudillo-Defru}, {Allart}, {Bonfils}, {S{\'e}gransan}, {Delfosse}, {Cegla}, {Wyttenbach}, {Heng}, {Lavie}, \& {Pepe}}]{Bourrier2018}
{Bourrier}, V., {Lovis}, C., {Beust}, H., {et~al.} 2018, \nat, 553, 477, \dodoi{10.1038/nature24677}

\bibitem[{{Bourrier} {et~al.}(2023){Bourrier}, {Attia}, {Mallonn}, {Marret}, {Lendl}, {Konig}, {Krenn}, {Cretignier}, {Allart}, {Henry}, {Bryant}, {Leleu}, {Nielsen}, {Hebrard}, {Hara}, {Ehrenreich}, {Seidel}, {dos Santos}, {Lovis}, {Bayliss}, {Cegla}, {Dumusque}, {Boisse}, {Boucher}, {Bouchy}, {Pepe}, {Lavie}, {Rey Cerda}, {S{\'e}gransan}, {Udry}, \& {Vrignaud}}]{Bourrier2023}
{Bourrier}, V., {Attia}, O., {Mallonn}, M., {et~al.} 2023, \aap, 669, A63, \dodoi{10.1051/0004-6361/202245004}

\bibitem[{{Castro-Gonz{\'a}lez} {et~al.}(2026){Castro-Gonz{\'a}lez}, {Bourrier}, {Ehrenreich}, {Armstrong}, {Correia}, \& {Lendl}}]{CastroGonzalez2026}
{Castro-Gonz{\'a}lez}, A., {Bourrier}, V., {Ehrenreich}, D., {et~al.} 2026, \aap, 709, L17, \dodoi{10.1051/0004-6361/202659558}

\bibitem[{{Castro-Gonz{\'a}lez} {et~al.}(2024){Castro-Gonz{\'a}lez}, {Bourrier}, {Lillo-Box}, {Delisle}, {Armstrong}, {Barrado}, \& {Correia}}]{CastroGonzalez2024}
{Castro-Gonz{\'a}lez}, A., {Bourrier}, V., {Lillo-Box}, J., {et~al.} 2024, \aap, 689, A250, \dodoi{10.1051/0004-6361/202450957}

\bibitem[{{Correia} {et~al.}(2014){Correia}, {Bou{\'e}}, {Laskar}, \& {Rodr{\'\i}guez}}]{correia2014}
{Correia}, A. C.~M., {Bou{\'e}}, G., {Laskar}, J., \& {Rodr{\'\i}guez}, A. 2014, \aap, 571, A50, \dodoi{10.1051/0004-6361/201424211}

\bibitem[{{Correia} {et~al.}(2020){Correia}, {Bourrier}, \& {Delisle}}]{Correia2020}
{Correia}, A.~C.~M., {Bourrier}, V., \& {Delisle}, J.-B. 2020, \aap, 635, A37, \dodoi{10.1051/0004-6361/201936967}

\bibitem[{{Correia} \& {Valente}(2022)}]{correia_valente2022}
{Correia}, A. C.~M., \& {Valente}, E. F.~S. 2022, Celestial Mechanics and Dynamical Astronomy, 134, 24, \dodoi{10.1007/s10569-022-10079-3}

\bibitem[{{Dobos} \& {Turner}(2015)}]{DobosTurner2015_exomoons}
{Dobos}, V., \& {Turner}, E.~L. 2015, \apj, 804, 41, \dodoi{10.1088/0004-637X/804/1/41}

\bibitem[{{Driscoll} \& {Barnes}(2015)}]{Driscoll2015}
{Driscoll}, P., \& {Barnes}, R. 2015, Astrobiology, 15, 739, \dodoi{10.1089/ast.2015.1325}

\bibitem[{{Espinoza-Retamal} {et~al.}(2024){Espinoza-Retamal}, {Stef{\'a}nsson}, {Petrovich}, {Brahm}, {Jord{\'a}n}, {Sedaghati}, {Lucero}, {Tala Pinto}, {Mu{\~n}oz}, {Boyle}, {Leiva}, \& {Suc}}]{Espinoza2024}
{Espinoza-Retamal}, J.~I., {Stef{\'a}nsson}, G., {Petrovich}, C., {et~al.} 2024, \aj, 168, 185, \dodoi{10.3847/1538-3881/ad70b8}

\bibitem[{{Espinoza-Retamal} {et~al.}(2026){Espinoza-Retamal}, {Winn}, {Brahm}, {Petrovich}, {Stef{\'a}nsson}, {Bhaskar}, {Koo}, {Jord{\'a}n}, {Tala Pinto}, {Hobson}, {Veldhuis}, {Rojas}, {Teske}, {Butler}, {Crane}, {Shectman}, {Vissapragada}, {Boyle}, {Leiva}, \& {Suc}}]{Espinoza2026}
{Espinoza-Retamal}, J.~I., {Winn}, J.~N., {Brahm}, R., {et~al.} 2026, arXiv e-prints, arXiv:2602.18553, \dodoi{10.48550/arXiv.2602.18553}

\bibitem[{{Hallatt} \& {Millholland}(2026{\natexlab{a}})}]{hallat2026a}
{Hallatt}, T., \& {Millholland}, S. 2026{\natexlab{a}}, \apj, 997, 139, \dodoi{10.3847/1538-4357/adfb75}

\bibitem[{{Hallatt} \& {Millholland}(2026{\natexlab{b}})}]{hallat2026b}
---. 2026{\natexlab{b}}, \apj, 997, 138, \dodoi{10.3847/1538-4357/ae129d}

\bibitem[{{Hansen}(2010)}]{Hansen2010}
{Hansen}, B. M.~S. 2010, \apj, 723, 285, \dodoi{10.1088/0004-637X/723/1/285}

\bibitem[{{Henning} {et~al.}(2009){Henning}, {O'Connell}, \& {Sasselov}}]{Henning2009}
{Henning}, W.~G., {O'Connell}, R.~J., \& {Sasselov}, D.~D. 2009, \apj, 707, 1000, \dodoi{10.1088/0004-637X/707/2/1000}

\bibitem[{{Herath} {et~al.}(2024){Herath}, {Boukar{\'e}}, \& {Cowan}}]{Herath2024}
{Herath}, M., {Boukar{\'e}}, C.-{\'E}., \& {Cowan}, N.~B. 2024, \mnras, 535, 2404, \dodoi{10.1093/mnras/stae2431}

\bibitem[{{Herath} {et~al.}(2026){Herath}, {Cowan}, {Boukar{\'e}}, \& {Dumberry}}]{Herath2026}
{Herath}, M., {Cowan}, N.~B., {Boukar{\'e}}, C.-{\'E}., \& {Dumberry}, M. 2026, arXiv e-prints, arXiv:2604.18682, \dodoi{10.48550/arXiv.2604.18682}

\bibitem[{{Hut}(1981)}]{Hut81}
{Hut}, P. 1981, \aap, 99, 126

\bibitem[{{Jackson} {et~al.}(2008){Jackson}, {Greenberg}, \& {Barnes}}]{jackson2008}
{Jackson}, B., {Greenberg}, R., \& {Barnes}, R. 2008, \apj, 678, 1396, \dodoi{10.1086/529187}

\bibitem[{{LoRusso} {et~al.}(2026){LoRusso}, {Petrovich}, \& {Gautham Bhaskar}}]{LoRusso2026}
{LoRusso}, R., {Petrovich}, C., \& {Gautham Bhaskar}, H. 2026, arXiv e-prints, arXiv:2604.08383, \dodoi{10.48550/arXiv.2604.08383}

\bibitem[{{Lu} {et~al.}(2025){Lu}, {An}, {Li}, {Millholland}, {Rice}, {Brandt}, \& {Brandt}}]{Lu2025}
{Lu}, T., {An}, Q., {Li}, G., {et~al.} 2025, \apj, 979, 218, \dodoi{10.3847/1538-4357/ad9b79}

\bibitem[{{Mazeh} {et~al.}(2016){Mazeh}, {Holczer}, \& {Faigler}}]{Mazeh2016}
{Mazeh}, T., {Holczer}, T., \& {Faigler}, S. 2016, \aap, 589, A75, \dodoi{10.1051/0004-6361/201528065}

\bibitem[{{Meyer} {et~al.}(2010){Meyer}, {Elkins-Tanton}, \& {Wisdom}}]{Meyer2010}
{Meyer}, J., {Elkins-Tanton}, L., \& {Wisdom}, J. 2010, \icarus, 208, 1, \dodoi{10.1016/j.icarus.2010.01.029}

\bibitem[{{Ojakangas} \& {Stevenson}(1986)}]{Ojakangas1986}
{Ojakangas}, G.~W., \& {Stevenson}, D.~J. 1986, \icarus, 66, 341, \dodoi{10.1016/0019-1035(86)90163-6}

\bibitem[{{Renaud} \& {Henning}(2018)}]{Renaud2018}
{Renaud}, J.~P., \& {Henning}, W.~G. 2018, \apj, 857, 98, \dodoi{10.3847/1538-4357/aab784}

\bibitem[{{Rubenzahl} {et~al.}(2024){Rubenzahl}, {Howard}, {Halverson}, {Petrovich}, {Angelo}, {Stef{\'a}nsson}, {Dai}, {Householder}, {Fulton}, {Gibson}, {Roy}, {Shaum}, {Isaacson}, {Brodheim}, {Deich}, {Hill}, {Holden}, {Huber}, {Laher}, {Lanclos}, {Payne}, {Petigura}, {Schwab}, {Walawender}, {Wang}, {Weiss}, {Winn}, \& {Wright}}]{Rubenzahl2024}
{Rubenzahl}, R.~A., {Howard}, A.~W., {Halverson}, S., {et~al.} 2024, \apjl, 971, L40, \dodoi{10.3847/2041-8213/ad6985}

\bibitem[{{Segatz} {et~al.}(1988){Segatz}, {Spohn}, {Ross}, \& {Schubert}}]{Segatz1988}
{Segatz}, M., {Spohn}, T., {Ross}, M.~N., \& {Schubert}, G. 1988, \icarus, 75, 187, \dodoi{10.1016/0019-1035(88)90001-2}

\bibitem[{{Stef{\`a}nsson} {et~al.}(2022){Stef{\`a}nsson}, {Mahadevan}, {Petrovich}, {Winn}, {Kanodia}, {Millholland}, {Maney}, {Ca{\~n}as}, {Wisniewski}, {Robertson}, {Ninan}, {Ford}, {Bender}, {Blake}, {Cegla}, {Cochran}, {Diddams}, {Dong}, {Endl}, {Fredrick}, {Halverson}, {Hearty}, {Hebb}, {Hirano}, {Lin}, {Logsdon}, {Lubar}, {McElwain}, {Metcalf}, {Monson}, {Rajagopal}, {Ramsey}, {Roy}, {Schwab}, {Schweiker}, {Terrien}, \& {Wright}}]{Gummi2022}
{Stef{\`a}nsson}, G., {Mahadevan}, S., {Petrovich}, C., {et~al.} 2022, \apjl, 931, L15, \dodoi{10.3847/2041-8213/ac6e3c}

\bibitem[{{Storch} \& {Lai}(2014)}]{StorchLai2014}
{Storch}, N.~I., \& {Lai}, D. 2014, \mnras, 438, 1526, \dodoi{10.1093/mnras/stt2292}

\bibitem[{{Tian} {et~al.}(2017){Tian}, {Wisdom}, \& {Elkins-Tanton}}]{Tian2017}
{Tian}, Z., {Wisdom}, J., \& {Elkins-Tanton}, L. 2017, \icarus, 281, 90, \dodoi{10.1016/j.icarus.2016.08.030}

\bibitem[{{Wang} {et~al.}(2026){Wang}, {Wang}, \& {Batygin}}]{Wang2026}
{Wang}, X.-Y., {Wang}, S., \& {Batygin}, K. 2026, arXiv e-prints, arXiv:2605.28719.
\newblock \doarXiv{2605.28719}

\bibitem[{{Yu} \& {Dai}(2024)}]{YuDai2025}
{Yu}, H., \& {Dai}, F. 2024, \apj, 972, 159, \dodoi{10.3847/1538-4357/ad5ffb}

\bibitem[{{Zahnle} {et~al.}(2015){Zahnle}, {Lupu}, {Dobrovolskis}, \& {Sleep}}]{Zahnle2015}
{Zahnle}, K.~J., {Lupu}, R., {Dobrovolskis}, A., \& {Sleep}, N.~H. 2015, Earth and Planetary Science Letters, 427, 74, \dodoi{10.1016/j.epsl.2015.06.058}

\end{thebibliography}

\end{document}